\documentclass[superscriptaddress, nofootinbib, preprintnumbers, notitlepage, floatfix, longbibliography]{revtex4-1}

\usepackage{amsmath}	
\usepackage{amsthm}		
\usepackage{amssymb}	
\usepackage{eufrak}
\usepackage{datetime}
\usepackage{graphicx}
\usepackage{color}
\usepackage{verbatim}
\usepackage{bm}
\usepackage[colorlinks=true, citecolor=midblue, linkcolor=midblue, urlcolor=midblue]{hyperref}
\usepackage{mciteplus}
\usepackage{epigraph}
\usepackage{array}
\usepackage{soul}

\usepackage{setspace}

\usepackage{fancyhdr}

\numberwithin{equation}{section}

\definecolor{grey}{rgb}{0.4,0.4,0.4}
\definecolor{dullmagenta}{rgb}{0.4,0,0.4}
\definecolor{darkblue}{rgb}{0,0,0.4}
\definecolor{midblue}{rgb}{0,0,0.7}
\definecolor{midred}{rgb}{0.5,0,0}
\definecolor{orange}{rgb}{1,0.5,0}
\definecolor{lightbrown}{rgb}{0.75,0.5,0.25}
\definecolor{tan}{cmyk}{0.14,0.42,0.56,0}
\definecolor{djunglegreen}{cmyk}{0.99,0,0.52,0}
\definecolor{lightgreen}{rgb}{0,1,0}
\definecolor{olivegreen}{cmyk}{0.64,0,0.95,0.40}
\definecolor{midgreen}{rgb}{0.0,0.675,0.0}
\definecolor{darkgreen}{rgb}{0,0.5,0}

\newcommand{\vs}{\vspace}
\newcommand{\hs}{\hspace}
\renewcommand{\.}{\hspace{0.5mm}}

\newcommand{\la}{\ensuremath{\leftarrow}}

\newcommand{\Brm}{\ensuremath{\mathrm{B}}}
\newcommand{\Crm}{\ensuremath{\mathrm{C}}}

\newcommand{\Erm}{\ensuremath{\mathrm{E}}}
\newcommand{\Frm}{\ensuremath{\mathrm{F}}}

\newcommand{\Hrm}{\ensuremath{\mathrm{H}}}

\newcommand{\Krm}{\ensuremath{\mathrm{K}}}
\newcommand{\Lrm}{\ensuremath{\mathrm{L}}}

\newcommand{\Prm}{\ensuremath{\mathrm{P}}}
\newcommand{\Qrm}{\ensuremath{\mathrm{Q}}}
\newcommand{\Rrm}{\ensuremath{\mathrm{R}}}
\newcommand{\Srm}{\ensuremath{\mathrm{S}}}

\newcommand{\Vrm}{\ensuremath{\mathrm{V}}}

\newcommand{\arm}{\ensuremath{\mathrm{a}}}
\newcommand{\brm}{\ensuremath{\mathrm{b}}}
\newcommand{\crm}{\ensuremath{\mathrm{c}}}
\newcommand{\drm}{\ensuremath{\mathrm{d}}}

\newcommand{\frm}{\ensuremath{\mathrm{f}}}
\newcommand{\grm}{\ensuremath{\mathrm{g}}}
\newcommand{\hrm}{\ensuremath{\mathrm{h}}}

\newcommand{\prm}{\ensuremath{\mathrm{p}}}

\newcommand{\srm}{\ensuremath{\mathrm{s}}}

\newcommand{\Ocal}{\ensuremath{\mathcal{O}}}

\renewcommand{\d}{\ensuremath{\mathrm{d}}}

\newcommand{\keV}{\ensuremath{\mathrm{keV}}}
\newcommand{\MeV}{\ensuremath{\mathrm{MeV}}}
\newcommand{\GeV}{\ensuremath{\mathrm{GeV}}}

\newcommand{\eg}{e.g.}
\newcommand{\ie}{i.e.}

\newcommand{\cf}{cf.}
\newcommand{\be}{\begin{equation}}
\newcommand{\ee}{\end{equation}}
\newcommand{\ba}{\begin{eqnarray}}
\newcommand{\ea}{\end{eqnarray}}

\smallskipamount

\settimeformat{ampmtime}

\usepackage{float}

\def\ga{\mathrel{\raise.3ex\hbox{$>$\kern-.75em\lower1ex\hbox{$\sim$}}}}
\def\la{\mathrel{\raise.3ex\hbox{$<$\kern-.75em\lower1ex\hbox{$\sim$}}}}

\def\Msun{M_\odot}

\def\fPBH{f_{\rm PBH}}
\def\MPBH{M_{\rm PBH}}

\def\dCP{\delta_{\rm CP}}
\def\nB{n_{\rm B}}
\def\OR{\Omega_{\rm R}}
\def\OM{\Omega_{\rm M}}
\def\OB{\Omega_{\rm B}}

\def\fPBH{f_{\rm PBH}}

\usepackage[width=16cm, height=24.2cm]{geometry}
\begin{document}

\title{Primordial Black Holes as Dark Matter: Recent Developments}

\author{Bernard Carr}
\email{b.j.carr@qmul.ac.uk}
\affiliation{School of Physics and Astronomy,
	Queen Mary University of London,
	Mile End Road,
	London E1 4NS,
	United Kingdom}
	
\author{Florian K{\"u}hnel}
\email{kuhnel@kth.se}
\affiliation{
	Arnold Sommerfeld Center,
	Ludwig-Maximilians-Universit{\"a}t,
	Theresienstra{\ss}e 37,
	80333 M{\"u}nchen,
	Germany}

\date{\formatdate{\day}{\month}{\year}, \currenttime}

\begin{abstract}
\vs{1mm}
\hs{3mm}\begin{minipage}[h][45mm][b]{0.8\textwidth}
Although the dark matter is usually assumed to be some form of elementary particle, primordial black holes (PBHs) could also provide some of it. However, various constraints restrict the possible mass windows to $10^{16}$ -- $10^{17}\,$g, $10^{20}$ -- $10^{24}\,$g and $10$ -- $10^{3}\,M_{\odot}$. The last possibility is contentious but of special interest in view of the recent detection of black-hole mergers by LIGO/Virgo. PBHs might have important consequences and resolve various cosmological conundra even if they have only a small fraction of the dark-matter density. In particular, those larger than $10^{3}\,M_{\odot}$ could generate cosmological structures through the {\it seed} or {\it Poisson} effect, thereby alleviating some problems associated with the standard cold dark-matter scenario, and sufficiently large PBHs might provide seeds for the supermassive black holes in galactic nuclei. More exotically, the Planck-mass relics of PBH evaporations or stupendously large black holes bigger than $10^{12}\,M_{\odot}$ could provide an interesting dark component.
\end{minipage}
\end{abstract}

\maketitle

\vfil
\begin{minipage}[h][-33mm][t]{0.877\textwidth}
\epigraph{\it \color{grey}Not from the stars do I my judgement pluck,\\
And yet methinks I have astronomy.\\
But not to tell of good or evil luck,\\
Of plagues, of dearths, or season's quality;\\
Nor can I fortune to brief minutes tell,\\
Or say with princes if it shall go well.}{\color{grey}William Shakespeare, {\it Sonnet 14}}
\end{minipage}

\newpage
\pagestyle{empty}
\tableofcontents
\clearpage

\section{Introduction}
\label{sec:Introduction}

\vs{-2mm}
\noindent Primordial black holes (PBHs) have been a source of interest for nearly 50 years \cite{1967SvA....10..602Z}, despite the fact that there is still no evidence for them. One reason for this interest is that only PBHs could be small enough for Hawking radiation to be important \cite{Hawking:1974rv}. This discovery has not yet been confirmed experimentally and there remain major conceptual puzzles associated with the process. Nevertheless, it is generally recognised as one of the key developments in 20th century physics because it beautifully unifies general relativity, quantum mechanics and thermodynamics. The fact that Hawking was only led to this discovery through contemplating the properties of PBHs illustrates that it can be useful to study something even if it does not exist! But, of course, the situation is much more interesting if PBHs do exist.

PBHs smaller than about $10^{15}\,$g would have evaporated by now with many interesting cosmological consequences \cite{Carr:2009jm}. Studies of such consequences have placed useful constraints on models of the early Universe and, more positively, evaporating PBHs have been invoked to explain certain features: for example, the extragalactic \cite{Page:1976wx, Carr:1976zz} and Galactic \cite{Wright:1995bi, Lehoucq:2009ge} $\gamma$-ray backgrounds, antimatter in cosmic rays \cite{Kiraly:1981ci, MacGibbon:1991vc}, the annihilation line radiation from the Galactic centre \cite{1980AA....81..263O, Bambi:2008kx}, the reionisation of the pregalactic medium \cite{Belotsky:2014twa} and some short-period $\gamma$-ray bursts \cite{1996MNRAS.283..626B, Cline:1996zg}. However, there are usually other possible explanations for these features, so there is no definitive evidence for evaporating PBHs. Only the original papers for each topic are cited here and a more comprehensive list of references can be found in Reference \cite{Carr:2009jm}. 

Attention has therefore shifted to the PBHs larger than $10^{15}\,$g, which are unaffected by Hawking radiation. Such PBHs might have various astrophysical consequences, such as providing the seeding of the supermassive black holes (SMBHs) in galactic nuclei \cite{1984MNRAS.206..801C, Bean:2002kx}, the generation of large-scale structure through Poisson fluctuations \cite{Meszaros:1975ef} and important effects on the thermal and ionisation history of the Universe \cite{1981MNRAS.194..639C}. Again only the original papers are cited here. But perhaps the most exciting possibility{\;---\;}and the main focus of this review{\;---\;}is that they could provide the dark matter (DM) which comprises $25\%$ of the critical density \cite{Carr:2016drx}, an idea that goes back to the earliest days of PBH research \cite{1975Natur.253..251C}. Since PBHs formed in the radiation-dominated era, they are not subject to the well-known big bang nucleosynthesis (BBN) constraint that baryons can have at most $5\%$ of the critical density \cite{Cyburt:2003fe}. They should therefore be classified as non-baryonic and behave like any other form of cold dark matter (CDM) \cite{Frampton:2015xza}. It is sometimes assumed that they must form before BBN, implying an upper limit of $10^{5}\,\Msun$, but the fraction of the Universe in PBHs at that time would be tiny, so the effect on BBN might only be small.

As with other CDM candidates, there is still no compelling evidence that PBHs provide the dark matter. However, there have been claims of evidence from dynamical and lensing effects. In particular, there was a flurry of excitement in 1997, when the microlensing results of massive compact halo objects (MACHOs) suggested that the dark matter could be compact objects of mass $0.5\,\Msun$ \cite{Alcock:1996yv}. Alternative microlensing candidates could be excluded and PBHs of this mass might naturally form at the quark-hadron phase transition at $10^{-5}\,$s \cite{Jedamzik:1998hc}. Subsequently, however, it was shown that such objects could comprise for only $20\%$ of the dark matter and indeed the entire mass range $10^{-7}\,\Msun$ to $10\,\Msun$ was later excluded from providing all of it \cite{Alcock:2000ph, Tisserand:2006zx}. In recent decades attention has focused on other mass ranges in which PBHs could have a significant density and numerous constraints allow only three possibilities: the asteroid mass range ($10^{16}$ -- $10^{17}$\,g), the sublunar mass range ($10^{20}$ -- $10^{26}$\,g) and the intermediate mass range ($10$ -- $10^{3}\,\Msun$).

We discuss the constraints on $f( M )$, the fraction of the halo in PBHs of mass $M$, in Section \ref{sec:Constraints-and-Caveats}; this is a much reduced version of the recent review by Carr {\it et al.}~\cite{Carr:2020gox}. The results are summarised in Figure \ref{fig:contraints-large}, all the limits assuming that the PBHs have a monochromatic mass function and cluster in the Galactic halo in the same way as other forms of CDM. Although, for the sake of completeness, we include evaporating PBHs, we do not focus on them in this review except insomuch as they may leave stable Planck-mass relics because these could also be dark-matter candidates. However, if Hawking evaporation were avoided for some reason, it is worth stressing that PBHs could provide the dark matter all the way down to the Planck mass, with few (if any) non-gravitational constraints.

At first sight, the implication of Figure \ref{fig:contraints-large} is that PBHs are excluded from having an appreciable density in almost every mass range. However, our intention is not to put nails in the coffin of the PBH scenario because every constraint is a potential signature. In particular, we have mentioned that there are still some mass windows in which PBHs could provide the dark matter. PBHs could be generated by inflation in all of these windows but theorists are split as to which one they favour. For example, Inomata {\it et al.}~\cite{Inomata:2017okj} argue that double inflation can produce a peak at around $10^{20}\,$g, while Clesse and Garc{\'i}a-Bellido \cite{Clesse:2015wea} argue that hybrid inflation can produce a peak at around $10\,\Msun$. A peak at the latter mass could also be produced by a reduction in the pressure at the quark-hadron phase transition \cite{Byrnes:2018clq}, even if the primordial fluctuations have no feature on that scale. There is a parallel here with the search for particle dark matter, where there is also a split between groups searching for light and heavy candidates.
\newpage

\pagestyle{fancy}
It should be stressed that non-evaporating PBHs are dark even if they do not provide {\it all} the dark matter, so this review does not focus exclusively on the proposal that PBHs solve the dark-matter problem. Many objects are dark, so it is not implausible that the dark matter comprises some mixture of PBHs and WIMPs. Indeed, we will see that this situation would have interesting consequences for both. Also, even if PBHs provide only a small fraction of the dark matter, they may still be of great cosmological interest. For example, they could play a r{\^o}le in generating the supermassive black holes in galactic nuclei and these have obvious astrophysical significance even though they provide only $0.1\%$ of the dark matter.

The constraints shown in Figure \ref{fig:contraints-large} assume that the PBH mass function is monochromatic (\ie~with a width $\Delta M \sim M$). However, there are many scenarios in which one would expect the mass function to be extended. For example, inflation often produces a lognormal mass function \cite{Dolgov:1992pu} and critical collapse generates an extended low mass tail \cite{Yokoyama:1998xd}. In the context of the dark-matter problem, this is a double-edged sword. On the one hand, it means that the {\it total} PBH density may suffice to explain the dark matter, even if the density in any particular mass band is small and within the observational bounds. On the other hand, even if PBHs can provide all the dark matter at some mass scale, the extended mass function may still violate the constraints at some other scale \cite{Green:2016xgy}. While there is now a well-understood procedure for analysing constraints in the extended case \cite{Carr:2017jsz}, identifying the optimal PBH mass window remains problematic \cite{Kuhnel:2017pwq}.

The proposal that the dark matter could comprise PBHs in the intermediate mass range has attracted much attention recently as a result of the LIGO/Virgo detections of merging binary black holes with mass in the range $10$ -- $50\,\Msun$ \cite{TheLIGOScientific:2016pea, TheLIGOScientific:2016wfe, LIGOScientific:2018jsj}. Since the black holes are larger than initially expected, it has been suggested that they could represent a new population, although the mainstream view remains that they are the remnants of ordinary stars \cite{2016Natur.534..512B}. One possibility is that they were of Population III origin (\ie~forming between decoupling and galaxy formation). Indeed, the suggestion that LIGO might detect gravitational waves from coalescing intermediate mass Population III black holes was first made more than 30 years ago \cite{1984MNRAS.207..585B} and, rather remarkably, Kinugawa {\it et al.}~predicted a Population III coalescence peak at $30\,\Msun$ shortly before the first LIGO detection of black holes of that mass \cite{Kinugawa:2014zha}. Another possibility, more relevant to the present considerations, is that the LIGO/Virgo black holes are primordial, as first discussed in Reference \cite{Nakamura:1997sm}. However, this does not require the PBHs to provide {\it all} the dark matter. While this possibility has been suggested \cite{Bird:2016dcv}, the predicted merger rate depends on when the binaries form and uncertain astrophysical factors, so the dark-matter fraction could still be small \cite{Sasaki:2016jop, Nakamura:2016hna, Sasaki:2018dmp}. Indeed, the LIGO/Virgo results have already been used to constrain the PBH dark-matter fraction \cite{Raidal:2017mfl}, although the limit is sensitive to the predicted merger rate, which is very model-dependent \cite{Ali-Haimoud:2017rtz}. Note that the PBH density should peak at a lower mass than the coalescence signal for an extended PBH mass function, since the gravitational-waves amplitude scales as the black-hole mass.

The plan of this review paper is as follows: In Section \ref{sec:Primordial-Black-Hole-Formation} we elaborate on several aspects of PBH formation, including a general discussion of their mass and density, a review of PBH formation scenarios, and a consideration of the effects of non-Gaussianity and non-sphericity. In Section \ref{sec:Constraints-and-Caveats} we review current constraints on the density of PBH with a monochromatic mass function, these being associated with a variety of lensing, dynamical, accretion and gravitational-wave effects. At first sight, these seem to exclude PBHs providing the dark matter in {\it any} mass range but this conclusion may be avoided for an extended mass function and most limits are subject to caveats anyway. More positively, in Section \ref{sec:Claimed-Signatures} we overview various observational conundra which can be explained by PBHs, especially those associated with intermediate mass and supermassive black holes. In Section \ref{sec:Unified-Primordial-Black-Hole-Scenario} we discuss how the thermal history of the Universe naturally provides peaks in the PBH mass function at the mass scales associated with these conundra, the bumpy mass function obviating some of the limits discussed in Section \ref{sec:Constraints-and-Caveats}. We also present a recently-developed mechanism which helps to resolve a long-standing fine-tuning problem associated with PBH formation. In Section \ref{sec:Primordial-Black-Hole-versus-Particle-Dark-Matter} we discuss scenarios which involve a mixture of PBHs and particle dark matter. In Section \ref{sec:Conclusion} we draw some general conclusions about PBHs as dark matter.\vs{-2mm}

\section{Primordial Black Hole Formation}
\label{sec:Primordial-Black-Hole-Formation}

\noindent PBHs could have been produced during the early Universe due to various mechanisms. For all of these, the increased cosmological energy density at early times plays a major r{\^o}le \cite{Hawking:1971ei, Carr:1974nx}, yielding a rough connection between the PBH mass and the horizon mass at formation:
\begin{align}
	M
		&\sim
					\frac{ c^{3}\, t }{ G }
		\sim
					10^{15}\mspace{-1mu}
					\left(
						\frac{ t }{ 10^{-23}\,\srm }
					\right)
					\grm
					\, .
					\label{eq:Moft}
\end{align}
Hence PBHs could span an enormous mass range: those formed at the Planck time ($10^{-43}\,\srm$) would have the Planck mass ($10^{-5}\,\grm$), whereas those formed at $1\,\srm$ would be as large as $10^{5}\,\Msun$, comparable to the mass of the holes thought to reside in galactic nuclei. By contrast, black holes forming at the present epoch (\eg~in the final stages of stellar evolution) could never be smaller than about $1\,\Msun$. In some circumstances PBHs may form over an extended period, corresponding to a wide range of masses. Even if they form at a single epoch, their mass spectrum could still extend much below the horizon mass due to ``critical phenomena'' \cite{Gundlach:1999cu, Musco:2012au, Gundlach:2002sx, Niemeyer:1997mt, Niemeyer:1999ak, Shibata:1999zs, Musco:2004ak, Musco:2008hv, Kuhnel:2015vtw}, although most of the PBH density would still be in the most massive ones.
\vs{-2mm}

\subsection{Mass and Density Fraction of Primordial Black Holes}
\label{sec:Mass-and-Density-Fraction-of-Primordial-Black-Holes}

\noindent The fraction of the mass of the Universe in PBHs on some mass scale $M$ is epoch-dependent but its value at the formation epoch of the PBHs is denoted by $\beta( M )$. For the standard $\Lambda$CDM model, in which the age of the Universe is $t_{0} = 13.8\,{\rm Gyr}$, the Hubble parameter is $h = 0.68$ \cite{Ade:2015lrj} and the time of photon decoupling is $t_{\rm dec} = 380\,{\rm kyr}$ \cite{Hinshaw:2008kr}. If the PBHs have a monochromatic mass function, the fraction of the Universe's mass in PBHs at their formation time $t_{i}$ is related to their number density $n_{\rm PBH}( t_{i} )$ by \cite{Carr:2009jm}
\begin{align}
	\beta( M )
		&\equiv
					\frac{ M\,n_{\rm PBH}( t_{i} ) }{ \rho( t_{i} ) }
		\approx
					7.98 \times 10^{-29}\,
					\gamma^{-1/2}
					\left(
						\frac{ g_{* i} }{ 106.75 }
					\right)^{\!1/4}
					\left(
						\frac{ M }{ \Msun }
					\right)^{\!3/2}
					\left(
						\frac{ n_{\rm PBH}( t_{0} ) }{ 1\,{\rm Gpc}^{-3} }
					\right)
					,
					\label{eq:betaf}
\end{align}
where $\rho( t_{i} )$ is the density at time $t_{i}$ and $\gamma$ is the ratio of the PBH mass to the horizon mass. $g_{* i}$ is the number of relativistic degrees of freedom at PBH formation, normalised to its value at $10^{-5}\,\srm$ since it does not increase much before that in the Standard Model and this is the period in which most PBHs are likely to form. 

The current density parameter for PBHs which have not yet evaporated is
\begin{align}
	\Omega_{\rm PBH}
		=
					\frac{ M\,n_{\rm PBH}( t_{0} ) }
					{ \rho_{\rm crit} }
		\approx
					\left(
						\frac{ \beta( M ) }{ 1.03 \times 10^{-8} }
					\right)\!\!\.
					\left(
						\frac{ h }{ 0.68 }
					\right)^{\!- 2}
					\gamma^{1 / 2}
					\left(
						\frac{ g_{* i} }{ 106.75 } 
					\right)^{\!- 1 / 4}
					\left(
						\frac{ M }{ \Msun } 
					\right)^{\!- 1 / 2}
					\, ,
					\label{eq:omega}
\end{align}
where $\rho_{\rm crit}$ is critical density. Equation \eqref{eq:omega} can be expressed in terms of the ratio of the current PBH mass density to the CDM density:
\begin{align}
	f
		\equiv
					\frac{\Omega_{ {\rm PBH}} }
					{ \Omega_{\rm CDM} }
		\approx
					3.8\,\Omega_{\rm PBH}
		\approx
					2.4\,\beta_{\rm eq}
					\, ,				
\label{eq:f}
\end{align}
where $\beta_{\rm eq}$ is the PBH mass fraction at matter-radiation equality and we use the most recent value $\Omega_{\rm CDM} = 0.26$ indicated by Planck \cite{Aghanim:2018eyx}. The ratio of the energy densities of matter and radiation (all relativistic species) at any time is 
\begin{align}
	\frac{ \OM }
	{ \OR }
		=
					\frac{ \OB +\Omega_{\rm CDM} }
					{ \OR }
		\approx
					\frac{ 1700 }{ g_{*}( z ) }\.
					\frac{ 1 + \chi }{ 1 + z }
					\, ,
\end{align}
where $\chi \equiv \Omega_{\rm CDM} /\OB \approx 5$ is the ratio of the dark matter and baryonic densities. At PBH formation, the fraction of domains that collapse is
\vs{-1mm}
\begin{align}
	\beta
		\equiv
					\fPBH\,\frac{ \chi\;\OB }{ \OR }
		\simeq
					\fPBH\,\frac{ \chi\;\eta }
					{ g_{*}( T ) }\.
					\frac{ 0.7\,\GeV }{ T }
					\, ,
\end{align}
where $\eta = n_{\Brm} / n_{\gamma} = 6 \times 10^{-10}$ is the observed baryon-to-photon ratio (\ie~the baryon asymmetry prior to $10^{-5}\,$s). As discussed in Section \ref{sec:Unified-Primordial-Black-Hole-Scenario}, this relationship suggests a scenario in which baryogenesis is linked with PBH formation, with the smallness of the $\eta$ reflecting the rarity of the Hubble domains that collapse~\cite{Carr:2019hud}. The collapse fraction can also be expressed as
\vs{-1mm}
\begin{align}
	\beta
		\approx
					0.5\,f_{\rm tot}
					\left[
						\chi\.\gamma^{- 1 / 2}\.
						\eta\.g_{*}^{1 / 4}
					\right]\!
					\left(
						\frac{ M }{ \Msun }
					\right)^{\!1 / 2}
					,
					\label{eq:beta2}
\end{align}
where $f_{\rm tot}$ is the total dark-matter fraction and the square-bracketed term has a value of order $10^{-9}$.
\vs{-2mm}

\subsection{Formation Scenarios}
\label{sec:Formation-Scenarios}

\noindent We now review the large number of scenarios which have been proposed for PBH formation and the associated PBH mass functions. We have seen that PBHs generally have a mass of order the horizon mass at formation, so one might expect a monochromatic mass function (\ie~with a width $\Delta M \sim M$). However, in some scenarios PBHs form over a prolonged period and therefore have an extended mass function (\eg~with its form of the mass function depending on the power spectrum of the primordial fluctuations). As discussed below, even PBHs formed at a single epoch may have an extended mass function.

\subsubsection{Primordial Inhomogeneities}
\label{sec:Primordial-Inhomogeneities}

\vs{-2mm}
\noindent The most natural possibility is that PBHs form from primordial density fluctuations. Overdense regions will then stop expanding some time after they enter the particle horizon and collapse against the pressure if they are larger than the Jeans mass. If the horizon-scale fluctuations have a Gaussian distribution with dispersion $\sigma$, one expects for the fraction of horizon patches collapsing to a black hole to be \cite{Carr:1975qj}
\begin{align}
	\beta
		&\approx
					{\rm Erfc}\!
					\left[
						\frac{ \delta_{\crm} }{ \sqrt{2\.}\.\sigma }
					\right]
					.
					\label{eq:beta}
\end{align}
Here `Erfc' is the complementary error function and $\delta_{\crm}$ is the density-contrast threshold for PBH formation. In a radiation-dominated era, a simple analytic argument \cite{Carr:1975qj} suggests $\delta_{\crm} \approx 1 / 3$, but more precise numerical \cite{Musco:2012au} and analytical \cite{Harada:2013epa} investigations suggest $\delta_{\crm} = 0.45$. Note that there is a distinction between the threshold value for the density and curvature fluctuation \cite{Musco:2018rwt} and one now has a good analytic understanding of these issues \cite{Escriva:2019phb, Escriva:2020tak}. The threshold is also sensitive to any non-Gaussianity \cite{Atal:2018neu, Kehagias:2019eil}, the shape of the perturbation profile \cite{Kuhnel:2016exn, Escriva:2019nsa, Germani:2018jgr} and the equation of state of the medium (a feature exploited in Reference \cite{Carr:2019kxo}).

\subsubsection{Collapse from Scale-Invariant Fluctuations}
\label{sec:Collapse-from-Scale--Invariant-Fluctuations}

\vs{-2mm}
\noindent If the PBHs form from scale-invariant fluctuations (\ie~with constant amplitude at the horizon epoch), their mass spectrum should have the power-law form \cite{Carr:1975qj}
\begin{align}
	\frac{ \d n }{ \d M }
		&\propto
					M^{-\alpha}
	\quad
	{\rm with}
	\quad 
	\alpha
		=
					\frac{ 2\.(1 + 2\.w ) }
					{ 1 + w }
					\; ,
					\label{eq:spectrum}
\end{align}
where $\gamma$ specifies the equation of state ($p = w\.\rho\.c^{2}$) at PBH formation. The exponent arises because the background density and PBH density have different redshift dependencies. At one time it was argued that the primordial fluctuations would be {\it expected} to be scale-invariant \cite{1970PhRvD...1.2726H, 1972MNRAS.160P...1Z} but this does not apply in the inflationary scenario. Nevertheless, one would still expect the above equations to apply if the PBHs form from cosmic loops because the collapse probability is then scale-invariant. If the PBHs contain a fraction $f_{\rm DM}$ of the dark matter, this implies that the fraction of the dark matter in PBHs of mass larger than $M$ is 
\begin{align}
	f( M )
		\approx
					f_{\rm DM}
					\left(
						\frac{ \,M_{\rm DM} }{ M }
					\right)^{\!\!\alpha - 2}
					\quad
					(
						M_{\rm min}
						<
						M
						<
						M_{\rm max}
					)
					\; ,
					\label{eq:dark}
\end{align}
where $2 < \alpha < 3$, and $M_{\rm DM} \approx\,M_{\rm min}$ is the mass scale which contains most of the dark matter. In a radiation-dominated era, the exponent in Equation \eqref{eq:dark} becomes $1 / 2$.
\vs{2mm}

\subsubsection{Collapse in a Matter-Dominated Era}
\label{sec:Collapse-in-a-Matter--Dominated-Era}

\vs{-2mm}
\noindent PBHs form more easily if the Universe becomes pressureless (\ie~matter-dominated) for some period. For example, this may arise at a phase transition in which the mass is channeled into non-relativistic particles \cite{Khlopov:1980mg, 1982SvA....26..391P} or due to slow reheating after inflation \cite{Khlopov:1985jw, Carr:1994ar}. In a related context, Hidalgo {\it et al.}~have recently studied \cite{Hidalgo:2017dfp} PBH formation in a dust-like scenario of an oscillating scalar field during an extended period of preheating. Since the value of $\alpha$ in the above analysis is $2$ for $\gamma = 0$, one might expect $\rho( M )$ to increase logarithmically with $M$. However, the analysis breaks down in this case because the Jeans length is much smaller than the particle horizon, so pressure is not the main inhibitor of collapse. Instead, collapse is prevented by deviations from spherical symmetry and the probability of PBH formation can be shown to be \cite{Khlopov:1980mg} 
\begin{align}
	\beta( M )
		&=
					0.02\,\delta_{\Hrm}( M )_{}^{5}
					\; .
\end{align}
This is in agreement with the recent analysis of Harada {\it et al.}~\cite{Harada:2016mhb} and leads to a mass function
\begin{align}
	\frac{\d n}{\d M}
		&\propto
					M^{-2}\.\delta_{\Hrm}( M )_{}^{5}
					\; .
\end{align}
The collapse fraction $\beta( M )$ is still small for $\delta_{\Hrm}( M ) \ll 1$ but much larger than the exponentially suppressed fraction in the radiation-dominated case. If the matter-dominated phase extends from $t_{1}$ to $t_{2}$, PBH formation is enhanced over the mass range 
\begin{align}
	M_{\rm min}
		&\sim
					M_{\Hrm}( t_{1} )
					<
					M
					<
					M_{\rm max}
		\sim
					M_{\Hrm}( t_{2} )\,\delta_{\Hrm}
					(
						M_{\rm max}
					)_{}^{3 / 2}
					\; .
\end{align} 
The lower limit is the horizon mass at the start of matter-dominance and the upper limit is the horizon mass when the regions which bind at the end of matter-dominance enter the horizon. This scenario has recently been studied in Reference \cite{Carr:2017edp}.

\subsubsection{Collapse from Inflationary Fluctuations}
\label{sec:Collapse-from-Inflationary-Fluctuations}

\vs{-2mm}
\noindent If the fluctuations generated by inflation have a blue spectrum (\ie~decrease with increasing scale) and the PBHs form from the high-$\sigma$ tail of the fluctuation distribution, then the exponential factor in Equation \eqref{eq:beta} might suggest that the PBH mass function should have an exponential upper cut-off at the horizon mass when inflation ends. This epoch corresponds to the reheat time $t_{\Rrm}$, which the cosmic microwave background (CMB) quadrupole anisotropy requires to exceed $10^{-35}\,$s, so this argument places a lower limit of around $1\,$g on the mass of such PBHs. The first inflationary scenarios for PBH formation were proposed in References \cite{Carr:1993aq, Ivanov:1994pa, GarciaBellido:1996qt, Randall:1995dj} and subsequently there have been a huge number of papers on this topic. In some scenarios, the PBHs form from a smooth symmetric peak in the inflationary power spectrum, in which case the PBH mass function should have the lognormal form:
\vs{-1mm}
\begin{align}
	\frac{ \d n }{ \d M }
		&\propto
					\frac{ 1 }{ M^{2} }\.
					\exp\!
					\left[
						-
						\frac{
							(
								\log M
								-
								\log M_{\crm}
							)^{2} }
						{ 2\.\sigma^{2} }
					\right]
					.
					\label{eq:mf}
\end{align}
This form was first suggested by Dolgov \& Silk \cite{Dolgov:1992pu} (see also References \cite{Dolgov:2008wu, Clesse:2015wea}) and has been demonstrated both numerically \cite{Green:2016xgy} and analytically \cite{Kannike:2017bxn} for the case in which the slow-roll approximation holds. It is therefore representative of a large class of inflationary scenarios, including the axion-curvaton and running-mass inflation models considered by K{\"u}hnel {\it et al.}~\cite{Kuhnel:2015vtw}. Equation \eqref{eq:mf} implies that the mass function is symmetric about its peak at $M_{\crm}$ and described by two parameters: the mass scale $M_{\crm}$ itself and the width of the distribution $\sigma$. The integrated mass function is
\begin{align}
	f( M )
		&=
					\int_{M}\d \tilde{M}\;
					\tilde{M}\.\frac{ \d n }{ \d \tilde{M} }
		\approx
					{\rm erfc}\!
					\left(
						\ln \frac{ M }{ \sigma }
					\right)
					.
\end{align}
However, not all inflationary scenarios produce the mass function \eqref{eq:mf}. Inomata {\it et al.}~\cite{Inomata:2016rbd} propose a scenario which combines a broad mass function at low $M$ (to explain the dark matter) with a sharp one at high mass (to explain the LIGO events).
\vs{-1mm}

\subsubsection{Quantum Diffusion}
\label{sec:Quantum-Diffusion}

\vs{-2mm}
\noindent Most of the relevant inflationary dynamics happens in regimes in which the classical inflaton-field evolution dominates over the field's quantum fluctuations. Under certain circumstances, however, the situation is reversed. There are two cases for which this happens. The first applies when the inflaton assumes larger values of its potential $\Vrm( \varphi )$, yielding eternally expanding patches of the Universe \cite{Vilenkin:1983xq, Starobinsky:1986fx, Linde:1986fd}. The second applies when the inflaton potential possesses one or more plateau-like features. Classically, using the slow-roll conditions, $| \overset{..}{\varphi} | \ll 3\.H\.| \overset{.}{\varphi} |$, $( \overset{.}{\varphi} )^{2} \ll 2\.\Vrm( \varphi )$, where an overdot represents a derivative with regard to cosmic time $t$, $H \equiv \overset{.}{a} / a$ is the Hubble parameter and $a$ is the scale factor, the number of inflationary e-folds is $N = \int\d\varphi\; H / \overset{.}{\varphi}$, which implies $\delta \varphi_{\Crm} = \overset{.}{\varphi} / H$. On the other hand, the corresponding quantum fluctuations are $\delta \varphi_{\Qrm} = H / 2 \pi$. Since the primordial metric perturbation is
\begin{align}
	\zeta
		&=
					\frac{ H }{ \overset{.}{\varphi} }\.
					\delta \varphi
		=
					\frac{ \delta \varphi_{\Qrm} }{ \delta \varphi_{\Crm} }
					\; ,
\end{align}
quantum effects are expected to be important whenever this quantity becomes of order one, \ie~$\zeta \sim \Ocal( 1 )$. This is often the case for PBH formation, where recent investigations indicate an increase of the power spectrum and hence the PBH abundance \cite{Pattison:2017mbe}. This quantum diffusion is inherently non-perturbative and so K{\"u}hnel \& Freese \cite{Kuhnel:2019xes} have developed a dedicated resummation technique in order to incorporate all higher-order corrections (see References \cite{Kuhnel:2008yk, Kuhnel:2008wr, Kuhnel:2010pp} for an application of these techniques to stochastic inflation). Ezquiaga {\it et al.}~have argued that quantum diffusion generically generates a high degree of non-Gaussianity \cite{Ezquiaga:2018gbw, Ezquiaga:2019ftu}.
\vs{-3mm}

\subsubsection{Critical Collapse}
\label{sec:Critical-Collapse}

\vs{-2mm}
\noindent It is well known that black-hole formation is associated with critical phenomena \cite{Choptuik:1992jv} and various authors have applied this feature in investigations of PBH formation \cite{Koike:1995jm, Niemeyer:1997mt, Evans:1994pj, Kuhnel:2015vtw}. The conclusion is that the mass function has an upper cut-off at around the horizon mass but there is also a low-mass tail \cite{Yokoyama:1998qw}. If we assume for simplicity that the density fluctuations have a monochromatic power spectrum on some mass scale $K$ and identify the amplitude of the density fluctuation when that scale crosses the horizon, $\delta$, as the control parameter, then the black-hole mass is \cite{Choptuik:1992jv} 
\begin{align}
	M
		&=
					K\.
					\big(
						\delta
						-
						\delta_{\crm}
					\big)^{\eta}
					\; .
					\label{eq:2}
\end{align}
Here $K$ can be identified with a mass $M_{\frm}$ of order the particle horizon mass, $\delta_{\crm}$ is the critical fluctuation required for PBH formation and the exponent $\eta$ has a universal value for a given equation of state. For $\gamma = 1 / 3$, one has $\delta_{\crm} \approx 0.4$ and $\eta \approx 0.35$. Although the scaling relation \eqref{eq:2} is expected to be valid only in the immediate neighborhood of $\delta_{\crm}$, most black holes should form from fluctuations with this value because the probability distribution function declines exponentially beyond $\delta = \delta_{\crm}$ if the fluctuations are blue. Hence it is sensible to calculate the expected PBH mass function using Equation \eqref{eq:2}. This allows us to estimate the mass function independently of the form of the probability distribution function of the primordial density fluctuations. A detailed calculation gives the mass function \cite{Yokoyama:1998xd}
\begin{align}
	\frac{ \d n }{ \d M }
		&\propto
					\!
					\left(
						\frac{ M }{ \xi\,M_{\frm} }
					\right)^{\!1 / \eta - 1}\,
					\exp\!
					\left[
						-
						(
							1
							-
							\eta
						)\!
						\left(
							\frac{ M }{ \eta\,M_{\frm} }
						\right)^{\!\!1 / \eta}\.
					\right]
					,
					\label{eq:initialMF}
\end{align}
where $\xi \equiv ( 1 - \eta / s )^{\eta}$, $s = \delta_{\crm} / \sigma$, $M_{\frm} = K$ and $\sigma$ is the dispersion of $\delta$. The above analysis depends on the assumption that the power spectrum of the primordial fluctuations is monochromatic. As shown by K{\"u}hnel {\it et al.}~\cite{Kuhnel:2015vtw} for a variety of inflationary models, when a realistic model for the power spectrum is used, the inclusion of critical collapse can lead to a significant shift, lowering and broadening of the PBH mass spectra{\;---\;}in some cases by several orders of magnitude.

\subsubsection{Collapse at the Quantum-Chromodynamics Phase Transition}
\label{sec:Collapse-at-QCD-Phase-Transition}

\vs{-2mm}
\noindent At one stage it was thought that the quantum-chromodynamics (QCD) phase transition at $10^{-5}\,$s might be first-order. This would mean that the quark-gluon plasma and hadron phases could coexist, with the cosmic expansion proceeding at constant temperature by converting the quark-gluon plasma to hadrons. The sound-speed would then vanish and the effective pressure would be reduced, significantly lowering the threshold $\delta_{\crm}$ for collapse. PBH production during a first-order QCD phase transitions was first suggested by Crawford \& Schramm \cite{Crawford:1982yz} and later revisited by Jedamzik \cite{Jedamzik:1996mr}. The amplification of density perturbations due to the vanishing of the speed of sound during the QCD transition was also considered by Schmid and colleagues \cite{Schmid:1998mx, Widerin:1998my}, while Cardall \& Fuller developed a semi-analytic approach for PBH production during the transition \cite{Cardall:1998ne}. It is now thought unlikely that the QCD transition is 1st order but one still expects some softening in the equation of state. Recently Byrnes {\it et al.}~\cite{Byrnes:2018clq} have discussed how this softening{\;--\;}when combined with critical phenomena and the exponential sensitivity of $\beta( M )$ to the equation of state{\;--\;}could produce a significant change in the mass function. The mass of a PBH forming at the QCD epoch is
\begin{align}
	M
		&=
					\frac{ \gamma\.\xi^{2} }{ g_{*}^{1/2} }\mspace{-1mu}
					\left(
						\frac{ 45 }{ 16 \pi^{3} }
					\right)^{\!\!1 / 2}
					\frac{ M_{\rm Pl}^{3} }{ m_{\prm}^{2} } 
		\approx
					0.9
					\left(
						\frac{ \gamma}{ 0.2 }
					\right)\!
					\left(
						\frac{ g_{*} }{ 10 }
					\right)^{\!- 1 / 2}\!
					\left(
						\frac{ \xi }{ 5 }
					\right)^{\!2}
					\Msun
					\, ,
					\label{eq:QCDmass}
\end{align}
where $M_{\rm Pl}$ is the Planck mass, $m_{\prm}$ is the proton mass, $g_{*}$ is normalised appropriately and $\xi \equiv m_{\prm} / ( k_{\Brm}\.T ) \approx 5$ is the ratio of the proton mass to the QCD phase-transition temperature, and $k_{\Brm}$ is the Boltzmann constant. This is necessarily close to the Chandrasekhar (Ch) mass:
\begin{align}
	M^{}_{\rm Ch}
		&=
					\frac{ \omega }{ \tilde{\mu}^{2} }\!
					\left(
						\frac{ 3\pi }{ 4 }
					\right)^{\!\! 1 / 2}
					\frac{ M_{\rm Pl}^{3} }{ m_{\prm}^{2} }
		\simeq
					5.6\,\tilde{\mu}^{-2}\,
					\Msun
					\, ,
\end{align}
where $\omega = 2.018$ is a constant that appears in the solution of the Lane-Emden equation and $\tilde{\mu}$ is the number of electrons per nucleon (1 for hydrogen, 2 for helium). The two masses are very close for the relevant parameter choices. Since all stars have a mass in the range $( 0.1$ -- $10 )\,M_{\rm Ch}$, this has the interesting consequence that dark and visible objects have comparable masses. From Equation \eqref{eq:beta2} it is also interesting that the collapse fraction at the QCD epoch is
\begin{equation}
	\beta
		\approx
					0.4\.
					f^{\rm tot}\.\chi\.\eta\.\xi
		\approx
					10\,\eta
					\, ,
					\label{beta3}
\end{equation}
where we have assumed $f^{\rm tot} \approx 1$ and $\chi \approx 5.5$ at the last step. This result is easily understood since one necessarily has $\rho_{\brm} / \rho_{\gamma} \sim \eta$ at the QCD epoch. We exploit this result in Section \ref{sec:Unified-Primordial-Black-Hole-Scenario} by suggesting that the collapse fraction determines the baryon-asymmetry.

\subsubsection{Collapse of Cosmic Loops}
\label{sec:Collapse-of-Cosmic-Loops}

\vs{-2mm}
\noindent In the cosmic string scenario, one expects some strings to self-intersect and form cosmic loops. A typical loop will be larger than its Schwarzschild radius by the factor $( G\.\mu )^{-1}$, where $\mu$ is the string mass per unit length. If strings play a r{\^o}le in generating large-scale structure, $G\.\mu$ must be of order $10^{-6}$. However, as discussed by many authors \cite{Hawking:1987bn, Polnarev:1988dh, Garriga:1993gj, Caldwell:1995fu, MacGibbon:1997pu, Jenkins:2020ctp}, there is always a small probability that a cosmic loop will get into a configuration in which every dimension lies within its Schwarzschild radius. This probability depends upon both $\mu$ and the string correlation scale. Note that the holes form with equal probability at every epoch, so they should have an extended mass spectrum with \cite{Hawking:1987bn} 
\begin{align}
	\beta
		&\sim
					\left(
						G\.\mu
					\right)^{2 x - 4}
					\, , 
\end{align}
where $x \equiv L / s$ is the ratio of the string length to the correlation scale. One expects $2 < x < 4$ and requires $G\.\mu < 10^{-7}$ to avoid overproduction of PBHs.

\subsubsection{Collapse through Bubble Collisions}
\label{sec:Collapse-through-Bubble-Collisions}

\vs{-2mm}
\noindent Bubbles of broken symmetry might arise at any spontaneously broken symmetry epoch and various people have suggested that PBHs could form as a result of bubble collisions \cite{Crawford:1982yz, Hawking:1982ga, Kodama:1982sf, Leach:2000ea, Moss:1994iq, Kitajima:2020kig}. However, this happens only if the bubble-formation rate per Hubble volume is finely tuned: if it is much larger than the Hubble rate, the entire Universe undergoes the phase transition immediately and there is not time to form black holes; if it is much less than the Hubble rate, the bubbles are very rare and never collide. The holes should have a mass of order the horizon mass at the phase transition, so PBHs forming at the GUT epoch would have a mass of $10^{3}\,$g, those forming at the electroweak unification epoch would have a mass of $10^{28}\,$g, and those forming at the QCD (quark-hadron) phase transition would have mass of around $1\,\Msun$. There could also be wormhole production at a 1st-order phase transition \cite{Kodama:1981gu, Maeda:1985bq}. The production of PBHs from bubble collisions at the end of first-order inflation has been studied extensively in References \cite{Khlopov:1998nm, Konoplich:1999qq, Khlopov:1999ys, Khlopov:2000js}.

\subsubsection{Collapse of Domain Walls}
\label{sec:Collapse-of-Domain-Walls}

\vs{-2mm}
\noindent The collapse of sufficiently large closed domain walls produced at a 2nd-order phase transition in the vacuum state of a scalar field, such as might be associated with inflation, could lead to PBH formation \cite{Dokuchaev:2004kr}. These PBHs would have a small mass for a thermal phase transition with the usual equilibrium conditions. However, they could be much larger if one invoked a non-equilibrium scenario \cite{Rubin:2001yw}. Indeed, they could span a wide range of masses, with a fractal structure of smaller PBHs clustered around larger ones \cite{Khlopov:1998nm, Konoplich:1999qq, Khlopov:1999ys, Khlopov:2000js}. Vilenkin and colleagues have argued that bubbles formed during inflation would (depending on their size) form either black holes or baby universes connected to our Universe by wormholes \cite{Garriga:2015fdk, Deng:2016vzb}. In this case, the PBH mass function would be very broad and extend to very high masses \cite{Deng:2017uwc, Liu:2019lul}.
\vs{4mm}

\subsection{Non-Gaussianity and Non-Sphericity}
\label{sec:Non--Gaussianity-and-Non--Sphericity}

\noindent As PBHs form from the extreme high-density tail of the spectrum of fluctuations, their abundance is acutely sensitive to non-Gaussianities in the density-perturbation profile \cite{Young:2013oia, Bugaev:2013vba}. For certain models{\;---\;}such as the hybrid waterfall or simple curvaton models \cite{Bugaev:2011wy, Bugaev:2011qt, Sasaki:2006kq}{\;---\;}it has even been shown that no truncation of non-Gaussian parameters can be made to the model without changing the estimated PBH abundance \cite{Young:2013oia}. However, non-Gaussianity induced PBH production can have serious consequences for the viability of PBH dark matter. PBHs produced with non-Gaussianity lead to isocurvature modes that could be detected in the CMB \cite{Young:2015kda, Tada:2015noa}. With the current Planck exclusion limits \cite{Ade:2015lrj}, this argument implies that the non-Gaussianity parameters $f_{\rm NL}$ and $g_{\rm NL}$ for a PBH-producing theory are both less than $\Ocal( 10^{-3} )$. For theories like the curvaton and hybrid inflation models \cite{Linde:1993cn, Clesse:2015wea}, this leads to the immediate exclusion of PBH dark matter, as the isocurvature effects would be too large.

Non-sphericity has not yet been subject to extensive numerical studies of the kind in Reference \cite{Musco:2012au} but non-zero ellipticity leads to possibly large effects on the PBH mass spectra as shown by Reference \cite{Kuhnel:2016exn}. Therein, the authors give an approximate analytical approximation for the collapse threshold, which will be larger than in the spherical case,
\vs{-2mm}
\begin{align}
	\delta_{\rm ec} / \delta_{\crm}
		&\simeq
					1
					+
					\kappa\.
					\left(
						\frac{ \sigma^{2} }
						{ \delta_{\crm}^{2} }
					\right)^{\! \tilde{\gamma}}
					\, ,
\end{align}
with $\delta_{\crm}$ being the threshold value for spherical collapse, $\sigma^{2}$ the amplitude of the density power spectrum at the given scale, $\kappa = 9 / \sqrt{10\.\pi\,}$ and $\tilde{\gamma} = 1 / 2$. Note that Reference \cite{Sheth:1999su} had already obtained this result for a limited class of cosmologies but this did not include the case of ellipsoidal collapse in a radiation-dominated model. A thorough numerical investigation is still needed to precisely determine the change of the threshold for fully relativistic non-spherical collapse. Note also that the effect due to non-sphericities is partly degenerate with that of non-Gaussianities \cite{Kuhnel:2016exn}.
\vs{-1mm}

\subsection{Multi-Spike Mass Functions}
\label{sec:Multi--Spike-Mass-Functions}

\noindent If PBHs are to explain phenomena on different mass scales, it is pertinent to consider the possibility that the PBH mass spectrum has several spikes. There are two known recent mechanisms for generating such spikes. The first has been proposed by Cai {\it et al.}~\cite{Cai:2018tuh}, who have discussed a new type of resonance effect which leads to prolific PBH formation. This arises because the sound-speed can oscillate in some inflationary scenarios, leading to parametric amplification of the curvature perturbation and hence a significant peak in the power spectrum of the density perturbations on some critical scale. The resonances are in narrow bands around certain harmonic frequencies with one of the peaks dominating. It turns out, one can easily get a peak of order unity. Although most PBHs form at the first peak, a small number will also form at subsequent peaks. The second mechanism for generating multi-spiked PBH mass spectra has recently been proposed by Carr \& K{\"uh}nel \cite{Carr:2018poi} and has been demonstrated for most of the well-studied models of PBH formation. This mechanism relies on the choice of non-Bunch-Davies vacua, leading to oscillatory features in the inflationary power spectrum, which in turn generates oscillations in the PBH mass function with exponentially enhanced spikes.

\section{Constraints and Caveats}
\label{sec:Constraints-and-Caveats}

\begin{figure}
	\vs{-0.5mm}
	\centering
	\includegraphics[scale=1.2]{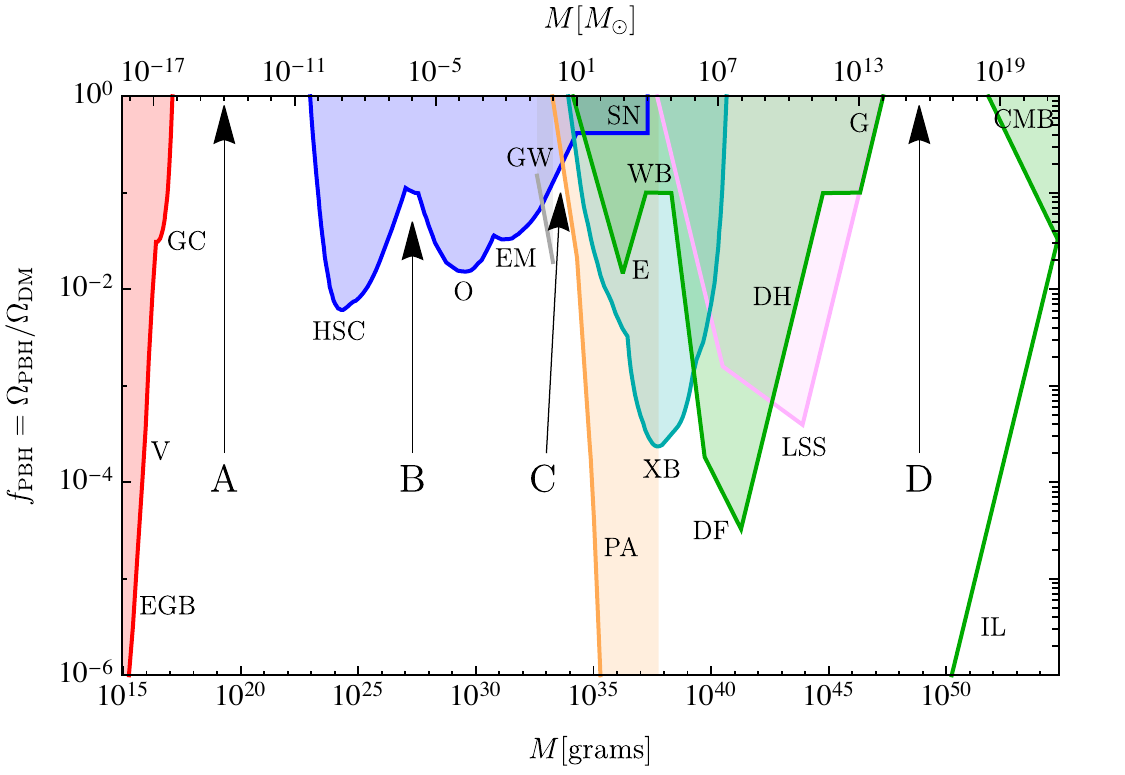}
	\vs{-0.5mm}
	\caption{
			Constraints on $f( M )$ 
			for a monochromatic 
			mass function, from 
			evaporations (red), 
			lensing (blue), 
			gravitational waves (GW) (gray),
			dynamical effects (green), 
			accretion (light blue), 
			CMB distortions (orange) and 
			large-scale structure (purple).
			Evaporation limits come from the extragalactic $\gamma$-ray background (EGB), 
			the Voyager positron flux (V) and 
			annihilation-line radiation from the Galactic centre (GC). 
			Lensing limits come from 
			microlensing of supernovae (SN) and of stars in M31 by Subaru (HSC), 
			the Magellanic Clouds by EROS and MACHO (EM) 
			and the Galactic bulge by OGLE (O). 
			Dynamical limits come from 
			wide binaries (WB), 
			star clusters in Eridanus II (E),
			halo dynamical friction (DF), 
			galaxy tidal distortions (G),
			heating of stars in the Galactic disk (DH) 
			and the CMB dipole (CMB). 
			Large-scale structure constraints derive from the requirement that 
			various cosmological structures do not form earlier than observed (LSS). 
			Accretion limits come from X-ray binaries (XB) and 
			Planck measurements of CMB distortions (PA).
			The incredulity limits (IL) correspond to one PBH per relevant environment 
			(galaxy, cluster, Universe). 
			There are four mass windows (A, B, C, D) 
			in which PBHs could have an appreciable density. 
			Possible constraints in window D are discussed in 
			Section \ref{sec:Primordial-Black-Hole-versus-Particle-Dark-Matter}
			but not in the past literature.\\[-3mm]}
	\label{fig:contraints-large}
\end{figure}

\begin{figure}
	\vs{1mm}
	\centering
	\includegraphics[width = 0.88 \textwidth]{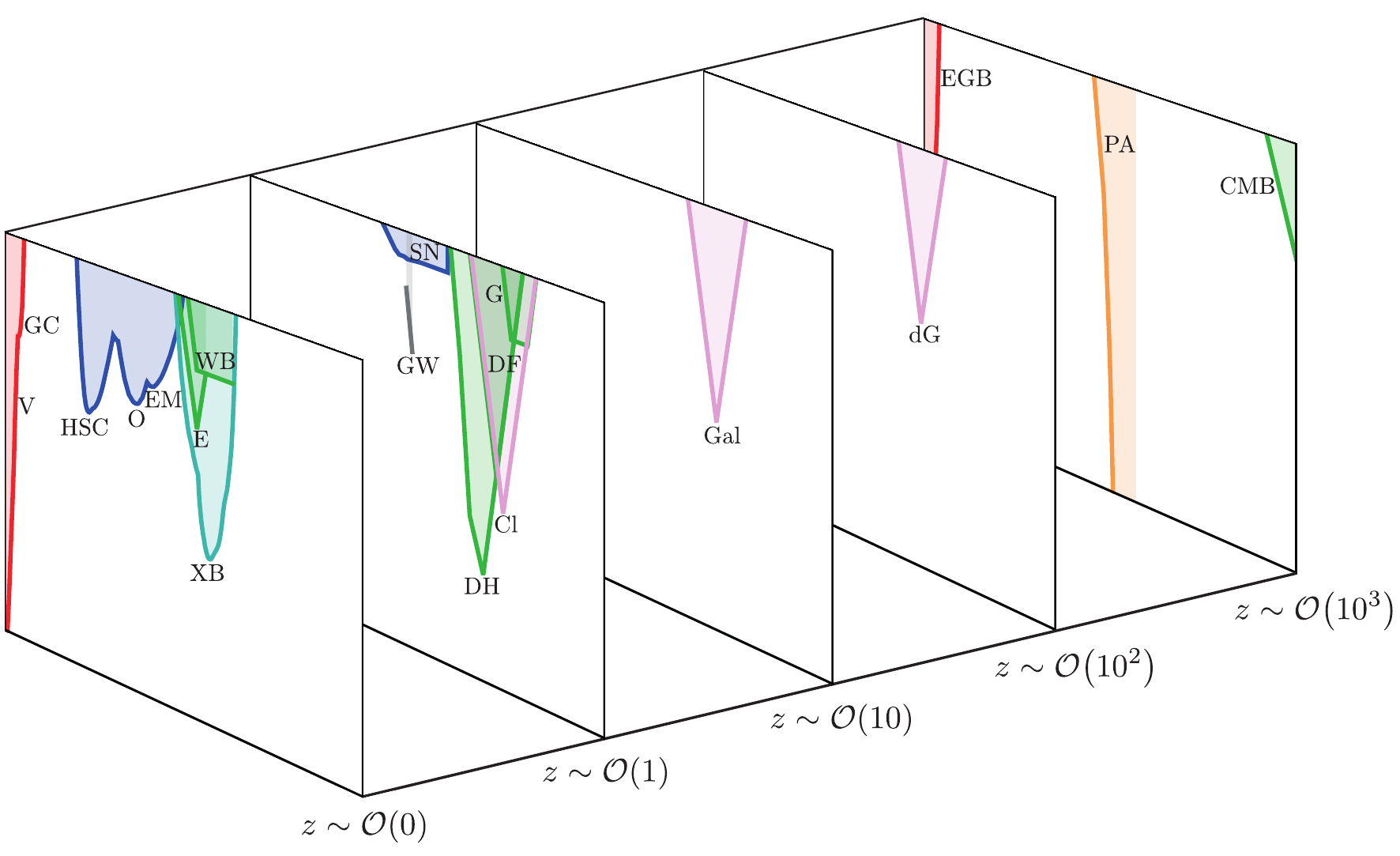}
	\caption{Sketch of the limits shown in Figure \ref{fig:contraints-large} 
			for different redshifts.
			Here, we break down
			the large-scale structure limit into its individual components from
				clusters (Cl),
				Milky Way galaxies (Gal) and
				dwarf galaxies (dG),
			as these originate from different redshifts (\cf~Reference~\cite{Carr:2018rid}).
			Further abbreviations are defined in the caption of 
			Figure \ref{fig:contraints-large}.}
			\vs{-3mm}
	\label{fig:PBH-constraints-for-different-Redshift}
\end{figure}

\noindent We now review the various constraints for PBHs which are too large to have evaporated completely by now, updating the equivalent discussion in References \cite{Carr:2009jm} and \cite{Carr:2016drx}. All the limits assume that PBHs cluster in the Galactic halo in the same way as other forms of CDM, unless they are so large that there is less than one per galaxy. Throughout this Section the PBHs are taken to have a monochromatic mass function, in the sense that they span a mass range $\Delta M \sim M$. In this case, the fraction $f( M )$ of the halo in PBHs is related to $\beta( M )$ by Equation \eqref{eq:omega}. Our limits on $f( M )$ are summarised in Figure \ref{fig:contraints-large}, which is based on Figure 10 of Reference \cite{Carr:2020gox}, this providing a much more comprehensive review of the PBH constraints. Following Reference \cite{Garcia-Bellido:2018leu}, the constraints are also broken down according to the redshift of the relevant observations in Figure \ref{fig:PBH-constraints-for-different-Redshift}. The main constraints derive from PBH evaporations, various gravitational-lensing experiments, numerous dynamical effects and PBH accretion. Where there are several limits in the same mass range, we usually show only the most stringent one. It must be stressed that the constraints in Figures \ref{fig:contraints-large} and \ref{fig:PBH-constraints-for-different-Redshift} have varying degrees of certainty and they all come with caveats. For some, the observations are well understood but there are uncertainties in the black-hole physics. For others, the observations themselves are not fully understood or depend upon additional astrophysical assumptions. The constraints may also depend on other physical parameters which are not shown explicitly. It is important to stress that some of the constraints can be circumvented if the PBHs have an extended mass function. Indeed, as discussed in Section \ref{sec:Unified-Primordial-Black-Hole-Scenario}, this may be {\it required} if PBHs are to provide most of the dark matter.

\subsection{Evaporation Constraints}
\label{sec:Evaporation-Constraints}

\noindent A PBH of initial mass $M$ will evaporate through the emission of Hawking radiation on a timescale $\tau \propto M^{3}$ which is less than the present age of the Universe for $M$ below $M_{*} \approx 5 \times 10^{14}$\,g \cite{Carr:2016hva}. There is a strong constraint on $f(M_{*} )$ from observations of the extragalactic $\gamma$-ray background \cite{Page:1976wx}. PBHs in the narrow band $M_{*} < M < 1.005\,M_{*}$ have not yet completed their evaporation but their current mass is below the mass $M_{q} \approx 0.4\,M_{*}$ at which quark and gluon jets are emitted. For $M > 2\,M_{*}$, one can neglect the change of mass altogether and the time-integrated spectrum of photons from each PBH is obtained by multiplying the instantaneous spectrum by the age of the Universe $t_{0}$. The instantaneous spectrum for primary (non-jet) photons is 
\begin{align}
	\frac{ \drm{\dot N}_\gamma^{\Prm} }{ \drm E }( M,\.E )
		&\propto
				\frac{ E^{2}\.\sigma( M,\.E ) }
				{ \exp( E M ) - 1 }
		\propto
				\begin{cases}
					E^{3}\.M^{3}
						& ( E < M^{-1} )
					\\[1mm]
					E^{2}\.M^{2}\,\exp( - E M )
						& ( E > M^{-1} )
					\; ,
				\end{cases}
\end{align}
where $\sigma( M,\.E )$ is the absorption cross-section for photons of energy $E$, so this gives an intensity
\vs{-1mm}
\begin{align}
	I( E )
		\propto
				f( M ) \times
				\begin{cases}
					E^{4}\.M^{2}
						& ( E < M^{-1} )
					\\[1mm]
					E^{3}\.M\,\exp( - E M )
						& ( E > M^{-1} )
					\; .
				\end{cases}
\end{align}
This peaks at $E^{\rm max} \propto M^{-1}$ with a value $I^{\rm max} ( M ) \propto f( M )\.M^{-2}$, whereas the observed intensity is $I^{\rm obs} \propto E^{- ( 1 + \epsilon )}$ with $\epsilon$ between $0.1$ and $0.4$, so putting $I^{\rm max}( M ) < I^{\rm obs}[ M( E ) ]$ gives \cite{Carr:2009jm}
\begin{align}
	f( M )
		<
				2 \times 10^{-8}
				\left(
					\frac{ M }{\,M_{*} }
				\right)^{\!\!3 + \epsilon}
		\quad
				( M > M_{*} )
				\; .
				\label{eq:photon2}
\end{align}
We plot this constraint in Figure \ref{fig:contraints-large} for $\epsilon = 0.2$. The Galactic $\gamma$-ray background constraint could give a stronger limit \cite{Carr:2016hva} but this depends sensitively on the form of the PBH mass function, so we do not discuss it here. 

There are various other evaporation constraints in this mass range. Boudad and Cirelli \cite{Boudaud:2018hqb} use positron data from Voyager 1 to constrain evaporating PBHs of mass $M < 10^{16}\,$g and obtain the bound $f < 0.001$. This complements the cosmological limit, as it is based on local Galactic measurements, and is also shown in Figure \ref{fig:contraints-large}. Laha \cite{Laha:2019ssq} and DeRocco and Graham \cite{DeRocco:2019fjq} constrain $10^{16}$ -- $10^{17}\,$g PBHs using measurements of the $511\,$keV annihilation line radiation from the Galactic centre. Other limits are associated with $\gamma$-ray and radio observations of the Galactic centre~\cite{Laha:2020ivk, Chan:2020zry} and the ionising effect of $10^{16}$ -- $10^{17}\,$g PBHs \cite{Belotsky:2014twa}.
\vs{-4mm}

\subsection{Lensing Constraints}
\label{sec:Lensing-Constraints}

\noindent Constraints on MACHOs with very low $M$ have been claimed from the femtolensing of $\gamma$-ray bursts (GRBs). Assuming the bursts are at a redshift $z \sim 1$, early studies implied $f < 1$ in the mass range $10^{-16}$ -- $10^{-13}\,\Msun$~\cite{Marani:1998sh, Nemiroff:2001bp} and $f < 0.1$ in the range $10^{-17}$ -- $10^{-14}\,\Msun$~\cite{Barnacka:2012bm}. However, Katz {\it et al.}~\cite{Katz:2018zrn} argue that most GRB sources are too large for these limits to apply, so we do not show them in Figure \ref{fig:contraints-large}. Kepler data from observations of Galactic sources \cite{Griest:2013esa, Griest:2013aaa} imply a limit in the planetary mass range: $f( M ) < 0.3$ for $2 \times 10^{-9}\,\Msun < M < 10^{-7}\,\Msun$. However, Niikura {\it et al.}~\cite{Niikura:2017zjd} have carried out a seven-hour observation of M31 with the Subaru Hyper Suprime-Cam (HSC) to search for microlensing of stars by PBHs lying in the halo regions of the Milky Way and M31 and obtain the much more stringent bound for $10^{-10} < M < 10^{-6}\,\Msun$ which is shown in Figure \ref{fig:contraints-large}.

Microlensing observations of stars in the Large and Small Magellanic Clouds probe the fraction of the Galactic halo in MACHOs in a certain mass range \cite{Paczynski:1985jf}. The optical depth of the halo towards LMC and SMC is related to the fraction $f( M )$ by $\tau^{({\rm SMC})}_{\mathrm L} = 1.4\,\tau^{({\rm LMC})}_{\mathrm L} = 6.6 \times 10^{-7}\,f( M )$ for the standard halo model \cite{Alcock:2000ph}. The MACHO project detected lenses with $M \sim 0.5\,\Msun$ but concluded that their halo contribution could be at most $10\%$ \cite{Hamadache:2006fw}, while the EROS project excluded $6 \times 10^{-8}\,\Msun < M < 15\,\Msun$ objects from dominating the halo. Since then, further limits in the range $0.1\,\Msun < M < 20\,\Msun$ have come from the OGLE experiment~\cite{Wyrzykowski:2009ep, CalchiNovati:2009kq, Wyrzykowski:2010bh, Wyrzykowski:2010mh, Wyrzykowski:2011tr}. The combined results can be approximated by
\begin{equation}
	f( M )
		<
					\begin{cases}
						1
							& ( 6 \times 10^{-8}\,\Msun < M < 30\,\Msun )
							\\[1mm]
						0.1
							& ( 10^{-6}\,\Msun < M < 1\,\Msun )
							\\[1mm]
						0.05
							& ( 10^{-3}\,\Msun < M < 0.4\,\Msun ) 
						\, .
					\end{cases}
\end{equation}
Recently Niikura {\it et al.}~\cite{Niikura:2019kqi} have used data from a five-year OGLE survey of the Galactic bulge to place much stronger limits in the range $10^{-6}\,\Msun< M < 10^{-4}\,\Msun$, although they also claim some positive detections. The precise form of the EROS and OGLE limits are shown in Figure \ref{fig:contraints-large}, while the possible detections are discussed in Section \ref{sec:Claimed-Signatures}.

PBHs cause most lines of sight to be demagnified relative to the mean, with a long tail of high magnifications. Zumalac{\'a}rregui and Seljak \cite{Zumalacarregui:2017qqd} have used the lack of lensing in type Ia supernovae (SNe) to constrain any PBH population, an approach that allows for the effects of large-scale structure and possible non-Gaussianity in the intrinsic SNe luminosity distribution. Using current JLA data, they derive a bound $f < 0.35$ for $10^{-2}\,\Msun< M < 10^{4}\,\Msun$, the finite size of SNe providing the lower limit, and this constraint is shown in Figure \ref{fig:contraints-large}. Garc{\'i}a-Bellido \& Clesse \cite{Garcia-Bellido:2017imq} argue that this limit can be weakened if the PBHs have an extended mass function or are clustered. There is some dispute about this but Figure \ref{fig:contraints-large} is only for a monochromatic mass function anyway.

The recent discovery of fast transient events in massive galaxy clusters is attributed to individual stars in giant arcs being highly magnified due to caustic crossing. Oguri {\it et al.}~\cite{Oguri:2017ock} argue that the particular event MACS J1149 excludes a high density of PBHs anywhere in the mass range $10^{-5}\,\Msun < M < 10^{2}\,\Msun$ because this would reduce the magnifications.
 
Early studies of the microlensing of quasars \cite{1994ApJ...424..550D} seemed to exclude the possibility of all the dark matter being in objects with $10^{-3}\,\Msun < M < 60\,\Msun$, although this limit preceded the $\Lambda$CDM picture. More recent studies of quasar microlensing suggest a limit \cite{2009ApJ...706.1451M} $f( M ) < 1$ for $10^{-3}\,\Msun< M < 60\,\Msun$, although we argue in Section \ref{sec:Claimed-Signatures} that these surveys may also provide positive evidence for PBHs. Millilensing of compact radio sources \cite{Wilkinson:2001vv} gives a limit 
\begin{equation}
	f( M )
		<
					\begin{cases}
						( M / 2 \times 10^{4}\,\Msun )^{-2}
							& ( M < 10^{5}\,\Msun )
							\\[1mm]
						0.06
							& ( 10^{5}\,\Msun < M < 10^{8}\,\Msun )
							\\[1mm]
						( M / 4 \times 10^{8}\,\Msun )^{2}
							& ( M > 10^{8}\,\Msun )
							\, .
					\end{cases}
\end{equation}
Although weaker than the dynamical constraints in this mass range, and not included this in Figure \ref{fig:contraints-large}, we mention it because it illustrates that lensing limits extend to very large values of $M$.

\subsection{Dynamical Constraints}
\label{sec:Dynamical-Constraints}

\noindent The effects of collisions of planetary-mass PBHs on astronomical objects have been a subject of long-standing interest, although we do not show these constraints in Figure \ref{fig:contraints-large} because they are controversial. Roncadelli {\it et al.}~\cite{Roncadelli:2009qj} have suggested that halo PBHs could be captured and swallowed by stars in the Galactic disc. The stars would eventually be accreted by the holes, producing radiation and a population of subsolar black holes which could only be of primordial origin and this leads to a constraint $f < ( M / 3 \times 10^{26}\,$g), corresponding to a \emph{lower} limit on the mass. Capela {\it et al.}~have constrained PBH dark matter by considering their capture by white dwarfs \cite{Capela:2012jz} or neutron stars \cite{Capela:2013yf}, while Pani and Loeb \cite{Pani:2014rca} have argued that this excludes PBHs from providing the dark matter throughout the sublunar window. However, these limits have been disputed \cite{Defillon:2014wla} because the dark-matter density in globular clusters is now known to be much lower values than assumed in these analyses~\cite{Ibata:2012eq}. Graham {\it et al.}~\cite{Graham:2015apa} argue that the transit of a PBH through a white dwarf (WD) causes localised heating through dynamical friction and initiates runaway thermonuclear fusion, causing the WD to explode as a supernova. They claim that the shape of the observed WD distribution excludes $10^{19}$ -- $10^{20}\,$g PBHs from providing the dark matter and that $10^{20}$ -- $10^{22}\,$g ones are constrained by the observed supernova rate. However, these limits are inconsistent with hydrodynamical simulations of Montero-Camacho {\it et al.}~\cite{Montero-Camacho:2019jte}, who conclude that this mass range is still allowed.

A variety of dynamical constraints come into play at higher mass scales \cite{Carr:1997cn}. Many of them involve the destruction of various astronomical objects by the passage of nearby PBHs. If the PBHs have density $\rho$ and velocity dispersion $v$, while the objects have mass $M_{\crm}$, radius $R_{\crm}$, velocity dispersion $v_{\crm}$ and survival time $t_{\Lrm}$, then the constraint has the form:
\begin{align}
	f( M ) <
				\begin{cases}
					M_{\crm}\.v / ( G\.M \rho\.t_{\Lrm}\.R_{\crm}) 
						& \big[ M < M_{\crm}( v / v_{\crm} ) \big]
						\\[1mm]
					M_{\crm} / ( \rho\.v_{\crm}\.t_{\Lrm}\.R_{\crm}^{2} )
						&
						\big[
							M_{\crm}(v / v_{\crm} ) < M < M_{\crm}
							( v / v_{\crm} )^{3}
						\big]
						\\[1mm]
					M\.v_{\crm}^{2} / 
					\big(
						\rho\.R_{\crm}^{2}\.v^{3}\.t_{\Lrm}
					\big)\.
					\exp\!
					\big[
						( M / M_{\crm} )
						( v_{\crm} / V )^{3}
					\big]
						&
						\big[
							M
							>
							M_{\crm}( v / v_{\crm} )^{3}
						\big] 
					\, .
				\end{cases}
				\label{eq:carsaklim}
\end{align}
The three limits correspond to disruption by multiple encounters, one-off encounters and non-impulsive encounters, respectively. The fraction is thus constrained over the mass range
\begin{align}
	\frac{ M_{\crm}\.v }
	{ G\.\rho_{\rm DM}\.t_{\Lrm}\.R_{\crm} }
		<
					M
		<
					M_{\crm}
					\left(
						\frac{ v }{ v_{\crm} }
					\right)^{\!3}
					\, ,
\end{align}
the limits corresponding to the values of $M$ for which $f = 1$. Various numerical factors are omitted in this discussion. These limits apply provided there is at least one PBH within the relevant environment, which is termed the `incredulity' limit \cite{Carr:1997cn}. For an environment of mass $M_{\Erm}$, this limit corresponds to the condition $f( M ) > ( M /\,M_{\Erm})$, where $M_{\Erm}$ is around $10^{12}\,\Msun$ for halos, $10^{14}\,\Msun$ for clusters and $10^{22}\,\Msun$ for the Universe. In some contexts the incredulity limit renders the third expression in Equation \eqref{eq:carsaklim} irrelevant.

One can apply this argument to wide binaries in the Galaxy, which are particularly vulnerable to disruption by PBHs~\cite{1985ApJ...290...15B, 1987ApJ...312..367W}. In the context of the original analysis of Reference \cite{Quinn:2009zg}, Equation \eqref{eq:carsaklim} gives a constraint $f( M ) < ( M / 500\,\Msun )^{-1}$ for $M < 10^{3}\,\Msun$, with $500\,\Msun$ representing the upper bound on the mass of PBHs which dominate the halo and $10^{3}\,\Msun$ being the mass at which the limit flattens off. Only the flat part of the constraint appears in Figure \ref{fig:contraints-large}. However, the upper limit has been reduced to $\sim 10\,\Msun$ in later work \cite{Monroy-Rodriguez:2014ula}, so the narrow window between the microlensing lower bound and the wide-binary upper bound is shrinking. On the other hand, Tian {\it et al.}~\cite{2020ApJS..246....4T} have recently studied more than 4000 halo wide binaries in the Gaia survey and detected a break in their separation distribution, possibly indicative of PBHs with $M > 10\,\Msun$.

A similar argument for the survival of globular clusters against tidal disruption by passing PBHs gives a limit $f( M ) < ( M / 3 \times 10^{4}\,\Msun )^{-1}$ for $M < 10^{6}\,\Msun$, although this depends sensitively on the mass and the radius of the cluster \cite{Carr:1997cn}. The upper limit of $3 \times 10^{4}\,\Msun$ is consistent with the numerical calculations of Moore \cite{1993ApJ...413L..93M}. In a related argument, Brandt \cite{Brandt:2016aco} infers an upper limit of $5\,\Msun$ from the fact that a star cluster near the centre of the dwarf galaxy Eridanus II has not been disrupted by halo objects. Koushiappas and Loeb \cite{Koushiappas:2017chw} have also studied the effects of black holes on the dynamical evolution of dwarf galaxies. They find that mass segregation leads to a depletion of stars in the centres of such galaxies and the appearance of a ring in the projected stellar surface density profile. Using Segue 1 as an example, they exclude the possibility of more than $4\%$ of the dark matter being PBHs of around $10\,\Msun$. One would also expect sufficiently large PBHs to disrupt Ultra-Faint Dwarf Galaxies and a recent study of $27$ UFDGs by Stegmann {\it et al.}~\cite{Stegmann:2019wyz} appears to exclude PBHs in the mass range $1$ -- $100\,\Msun$ from providing the dark matter. Only the Eridanus limit is shown in Figure \ref{fig:contraints-large}.

Halo objects will overheat the stars in the Galactic disc unless one has $f( M ) < ( M / 3 \times 10^{6}\,\Msun )^{-1}$ for $M < 3 \times 10^{9}\,\Msun$ \cite{1985ApJ...299..633L}. The incredulity limit, $f( M ) < ( M / 10^{12}\,\Msun )$ (corresponding one PBH per halo), takes over for $M > 3 \times 10^{9}\,\Msun$ and this is the only part appearing in Figure \ref{fig:contraints-large}. Another limit in this mass range arises because halo objects will be dragged into the nucleus of the Galaxy by the dynamical friction of various stellar populations, and this process leads to excessive nuclear mass unless $f( M )$ is constrained \cite{Carr:1997cn}. As shown in Figure \ref{fig:contraints-large}, this limit has a rather complicated form because there are different sources of friction and it also depends on parameters such as the halo core radius, but it bottoms out at $M \sim 10^{7}\,\Msun$ with a value $f \sim 10^{-5}$. 

There are also interesting limits for black holes which are too large to reside in galactic halos. The survival of galaxies in clusters against tidal disruption by giant cluster PBHs gives a limit $f( M ) < ( M / 7 \times 10^{9}\,\Msun )^{-1}$ for $M < 10^{11}\,\Msun$, with the limit flattening off for $10^{11}\,\Msun < M < 10^{13}\,\Msun$ and then rising as $f( M ) < M / 10^{14}\,\Msun$ due to the incredulity limit. This constraint is shown in Figure \ref{fig:contraints-large} with typical values for the mass and the radius of the cluster. If there were a population of huge intergalactic (IG) PBHs with density parameter $\Omega_{\rm IG}( M )$, each galaxy would have a peculiar velocity due to its gravitational interaction with the nearest one \cite{etde_6856669}. The typical distance to the nearest one should be $d \approx 30\,\Omega_{\rm IG}( M )^{-1/3} ( M / 10^{16}\,\Msun )^{1/3}\,{\rm Mpc}$, so this should induce a peculiar velocity $v_{\rm pec} \approx G M\.t_{0} / d^{2}$ over the age of the Universe. Since the CMB dipole anisotropy shows that the peculiar velocity of our Galaxy is only $400\,{\rm km}\,\srm^{-1}$, one infers $\Omega_{\rm IG} < ( M / 5 \times 10^{15}\,\Msun )^{-1/2}$, so this gives the limit on the far right of Figure \ref{fig:contraints-large}. This intersects the incredulity limit (corresponding to one PBH within the particle horizon) at $M \sim 10^{21}\,\Msun$.

Carr and Silk \cite{Carr:2018rid} point out that large PBHs could generate cosmic structures through the `seed' or `Poisson' effect even if $f$ is small. If a region of mass $\bar{M}$ contains PBHs of mass $M$, the initial fluctuation is $M / \bar{M}$ for the seed effect and $(f\.M / \bar{M} )^{1/2}$ for the Poisson effect, the fluctuation growing as $z^{-1}$ from the redshift of CDM domination ($z_{\rm eq} \approx 4000$). Even if PBHs do not play a r{\^o}le in generating cosmic structures, one can place interesting upper limits of the fraction of dark matter in them by requiring that various types of structure do not form too early. For example, if we apply this argument to Milky-Way-type galaxies, assuming these have a typical mass of $10^{12}\,\Msun$ and must not bind before a redshift $z_{\Brm} \sim 3$, we obtain
\begin{equation}
	f( M )
		<
					\begin{cases}
						( M / 10^{6}\,\Msun )^{-1}
							& ( 10^{6}\,\Msun < M \lesssim 10^{9}\,\Msun )
							\\[1mm]
						M/10^{12}\,\Msun
							& ( 10^{9}\,\Msun \lesssim M < 10^{12}\,\Msun )
							\, ,
					\end{cases}
					\label{eq:galaxy}
\end{equation}
with the second expression corresponding to having one PBH per galaxy. This limit bottoms out at $M \sim 10^{9}\,\Msun$ with a value $f \sim 10^{-3}$. Similar constraints apply for the first bound clouds, dwarf galaxies and clusters of galaxies and the limits for all the systems are collected together in Figure \ref{fig:contraints-large}. The Poisson effect also influences the distribution of the Lyman-alpha forest \cite{Afshordi:2003zb, Murgia:2019duy}; the associated PBH constraint has a similar form to Equation \eqref{eq:galaxy} but could be much stronger, with the lower limit of $10^{6}\,\Msun$ being reduced to around $10^{2}\,\Msun$.
\vs{2mm}

\subsection{Accretion Constraints}
\label{sec:Accretion-Constraint}

\noindent PBHs could have a large luminosity at early times due to accretion of background gas and this effect imposes strong constraints on their number density. However, the analysis of the problem is complicated because the black hole luminosity will generally boost the matter temperature of the background Universe well above the standard Friedmann value even if the PBH density is small, thereby reducing the accretion. Thus there are two distinct but related PBH constraints: one associated with the effects on the Universe's thermal history and the other with the generation of background radiation. This problem was first studied in Reference \cite{1981MNRAS.194..639C} and we briefly review that analysis here. Even though this study was incomplete and later superseded by more detailed numerical investigations, we discuss it because it is the only analysis which applies for very large PBHs.

Reference \cite{1981MNRAS.194..639C} assumes that each PBH accretes at the Bondi rate~\cite{Bondi:1952ni}
\begin{equation}
	\dot{M}
		\approx
					10^{11}
					\left(
						M / \Msun
					\right)^{2}\!
					\left( n / {\rm cm}^{-3}
					\right)\!
					\left(
						T / 10^{4}\,\Krm
					\right)^{-3/2}\mspace{1mu}
					{\rm g}\,{\srm}^{-1}
					\, ,
\end{equation}
where a dot indicates differentiation with respect to cosmic time $t$ and the appropriate values of $n$ and $T$ are those which pertain at the black-hole accretion radius:
\begin{equation}
	R_{\arm}
		\approx
					10^{14}\,( M / \Msun )
					\left(
						T / 10^{4}\,\Krm
					\right)^{-1}\,
					{\rm cm}
					\, .
\end{equation}
Each PBH will initially be surrounded by an HII region of radius $R_{\srm}$, where the temperature is slightly below $10^{4}\,$K and determined by the balance between photoionisation heating and inverse Compton cooling from the CMB photons. If $R_{\arm} > R_{\srm}$ or if the whole Universe is ionised (so that the individual HII regions have merged), the appropriate values of $n$ and $T$ are those in the background Universe ($\bar{n}$ and $\bar{T}$). In this case, after decoupling, $\dot{M}$ is epoch-independent so long as $\bar{T}$ has its usual Friedmann behaviour ($\bar{T} \propto z^{2}$). However, $\dot{M}$ decreases if $\bar{T}$ is boosted above the Friedmann value. If the individual HII regions have not merged and $R_{\arm} < R_{\srm}$, the appropriate values for $n$ and $T$ are those within the HII region. In this case, $T$ is close to $10^{4}\,$K and pressure balance at the edge of the region implies $n \sim \bar{n}\.( \bar{T} / 10^{4}\,\Krm )$. Thus $\dot{M} \propto z^{5}$ and rapidly decreases until $\bar{T}$ deviates from the standard Friedmann behaviour.

If the accreted mass is converted into outgoing radiation with efficiency $\epsilon$, the associated luminosity is
\begin{equation}
	L
		=
					\epsilon\.\dot{M}\.c^{2}
					\, .
\end{equation}
Reference \cite{1981MNRAS.194..639C} assumes that both $\epsilon$ and the spectrum of emergent radiation are constant. If the spectrum extends up to energy $E_{\rm max} = 10\.\eta\,$keV, the high-energy photons escape from the individual HII regions unimpeded, so most of the black-hole luminosity goes into background radiation or global heating of the Universe through photoionisation when the background ionisation is low and Compton scattering off electrons when it is high. Reference \cite{1981MNRAS.194..639C} also assumes that $L$ cannot exceed the Eddington luminosity,
\begin{equation}
	L_{\rm ED}
		= 
					4 \pi\,G M\.m_{\prm} / \sigma_{T}
		\approx
					10^{38}\.( M / \Msun )\,{\rm erg}\,\srm^{-1}
					\, ,
\end{equation}
and it is shown that a PBH will radiate at this limit for some period after decoupling providing
\begin{equation}
	M
		>
					M_{\rm ED}
		\approx
					10^{3}\.
					\epsilon^{-1}\.
					\Omega_{\grm}^{-1}
					\, ,
\end{equation}
where $\Omega_{\grm}$ is the gas density parameter. The Eddington phase persists until a time $t_{\rm ED}$ which depends upon $M$ and $\Omega_{\rm PBH}$ and can be very late for large values of these parameters.

The effect on the thermal history of the Universe is then determined for different ($\Omega_{\rm PBH},\,M$) domains. One has various possible behaviours: (1) $\bar{T}$ is boosted above $10^{4}\,$K, with the Universe being reionised, and possibly up to the temperature of the hottest accretion-generated photons; (2) $\bar{T}$ is boosted to $10^{4}\,$K but not above it because of the cooling of the CMB; (3) $\bar{T}$ does not reach $10^{4}\,$K, so the Universe is not re-ionised, but there is a period in which it increases; (4) $\bar{T}$ never increases but follows the CMB temperature, falling like $z$ rather than $z^{2}$, for a while; (5) $\bar{T}$ never deviates from Friedmann behaviour.

Constraints on the PBH density in each domain are derived by comparing the time-integrated emission from the PBHs with the observed background intensity in the appropriate waveband \cite{1979MNRAS.189..123C}. For example, in domain (1) the biggest contribution to the background radiation comes from the end of the Eddington phase and the radiation would currently reside in the $0.1$ -- $1\,$keV range, where $\Omega_{\Rrm} \sim 10^{-7}$. The associated limit on the PBH density parameter is then shown to be \cite{1979MNRAS.189..123C}
\begin{equation}
	\Omega_{\rm PBH}
		<
					( 10\.\epsilon )^{-5/6}\.
					( M / 10^{4}\,\Msun )^{-5/6}\.
					\eta^{5/4}\,\Omega_{\rm g}^{-5/6}
					\, .
					\label{eq:lightlimit}
\end{equation}
This limit does not apply if the PBH increases its mass appreciably as a result of accretion. During the Eddington phase, each black hole doubles its mass on the Salpeter timescale, $t_{\Srm} \approx 4 \times 10^{8}\,\epsilon\,$y \cite{Salpeter:1964kb}, so one expects the mass to increase by a factor $\exp ( t_{\rm ED}/t_{\Srm})$ if $t_{\rm ED} > t_{\Srm}$. Since most of the final black-hole mass generates radiation with efficiency $\epsilon$, the current energy of the radiation produced is $E( M ) \approx \epsilon\.M c^{2} / z( M )$ where $z( M )$ is the redshift at which most of the radiation is emitted. This must be less than $z_{\Srm} \approx 10\.\epsilon^{-2/3}$, the redshift at which the age of the Universe is comparable to $t_{\Srm}$. The current background-radiation density is therefore $\Omega_{\Rrm} \gtrsim z_{\Srm}\,\epsilon\,\Omega_{\rm PBH}$, so the constraint becomes
\begin{equation}
	\Omega_{\rm PBH}
		<
					\epsilon^{-1}\.z_{\Srm}\.\Omega_{\Rrm}
		\approx
					10^{-5}\.( 10\.\epsilon )^{-5/3}
					\,.
					\label{eq:accretion2}
\end{equation} 
This relates to the well-known Soltan constraint \cite{Soltan:1982vf} on the growth of the SMBHs that power quasars. The limit given by Equation \eqref{eq:lightlimit} therefore flattens off at large values of $M$.
\newpage

One problem with the above analysis is that the steady-state Bondi formula fails if the accretion timescale,
\vs{-2mm}
\begin{equation}
	t_{\Brm}
		\approx
					10^{5}\.( M / 10^{4}\,\Msun )
					\left(
						\frac{ T }{ 10^{4}\,\Krm }
					\right)^{\!-3/2}{\rm years}
					\; ,
\label{eq:steady}
\end{equation}
exceeds the cosmic expansion time, with the solution being described by self-similar infall instead. For $M > 10^{4}\,\Msun$, this applies at decoupling and so one has to wait until the time given by Equation \eqref{eq:steady} for the Bondi formula to apply. Therefore the above analysis applies only if most of the radiation is generated after this time. Otherwise the background light limit is weakened.

Later an improved analysis was provided by Ricotti and colleagues \cite{Mack:2006gz, Ricotti:2007au, Ricotti:2007jk}. They used a more realistic model for the efficiency parameter $\epsilon$, allowed for the increased density in the dark halo expected to form around each PBH and included the effect of the velocity dispersion of the PBHs on the accretion in the period after cosmic structures start to form. They found much stronger accretion limits by considering the effects of the emitted radiation on the spectrum and anisotropies of the CMB rather than the background radiation itself. Using FIRAS data to constrain the first, they obtained a limit $f( M ) < ( M / 1\,\Msun )^{-2}$ for $1\,\Msun < M \lesssim 10^{3}\,\Msun$; using WMAP data to constrain the second, they obtained a limit $f( M ) < ( M / 30\,\Msun )^{-2}$ for $30\,\Msun < M \lesssim 10^{4}\,\Msun$. The constraints flatten off above the indicated masses but are taken to extend up to $10^{8}\,\Msun$. Although these limits appeared to exclude $f = 1$ down to masses as low as $1\,\Msun$, they were very model-dependent and there was also a technical error (an incorrect power of redshift) in the calculation.

This problem has been reconsidered by several groups, who argue that the limits are weaker than indicated in Reference \cite{Ricotti:2007au}. Ali-Ha{\"i}moud and Kamionkowski \cite{Ali-Haimoud:2016mbv} calculate the accretion on the assumption that it is suppressed by Compton drag and Compton cooling from CMB photons and allowing for the PBH velocity relative to the background gas. They find the spectral distortions are too small to be detected, while the anisotropy constraints only exclude $f = 1$ above $10^{2}\,\Msun$. Horowitz \cite{Horowitz:2016lib} performs a similar analysis and gets an upper limit of $30\,\Msun$. Neither of these analyses includes the super-Eddington effects expected above some mass, and this should lead to a flattening of the constraint. Poulin {\it et al.}~\cite{Poulin:2017bwe, Serpico:2020ehh} argue that the spherical accretion approximation probably breaks down, with an accretion disk forming instead, and this affects the statistical properties of the CMB anisotropies. Provided the disks form early, these constraints exclude a monochromatic distribution of PBH with masses above $2\,\Msun$ as the dominant form of dark matter. Since this is the strongest accretion constraint, it is the only one shown in Figure \ref{fig:contraints-large}.

More direct constraints can be obtained by considering the emission of PBHs at the present epoch. For example, Gaggero {\it et al.}~\cite{Gaggero:2016dpq} model the accretion of gas onto a population of massive PBHs in the Milky Way and compare the predicted radio and X-ray emission with observational data. The possibility that $\Ocal( 10 )\,\Msun$ PBHs can provide all of the dark matter is excluded at $5 \sigma$ level by a comparison with the VLA radio catalog and the Chandra X-ray catalog. Similar arguments have been made by Manshanden {\it et al.}~\cite{Manshanden:2018tze}. PBH interactions with the interstellar medium should result in a significant X-ray flux, contributing to the observed number density of compact X-ray objects in galaxies. Inoue \& Kusenko and Lu {et al.}~\cite{Inoue:2017csr, Lu:2020bmd} use the data to constrain the PBH number density in the mass range from a few to $2 \times 10^{7}\,\Msun$ and their limit is shown in Figure \ref{fig:contraints-large}. However, De Luca {\it et al.}~\cite{DeLuca:2020fpg} have stressed that the change in the mass of PBHs due to accretion may modify the interpretation of the observational bounds on $f( M )$ at the present epoch. In the mass range $10$ -- $100\,\Msun$ this can raise existing upper limits by several orders of magnitude.

\subsection{Cosmic Microwave Background Constraints}
\label{sec:CMB-Constraints}

\noindent If PBHs form from the high-$\sigma$ tail of Gaussian density fluctuations, as in the simplest scenario \cite{Carr:1975qj}, then another interesting limit comes from the dissipation of these density fluctuations by Silk damping at a much later time. This process leads to a $\mu$-distortion in the CMB spectrum \cite{Chluba:2012we} for $7 \times 10^{6} < t/{\rm s} < 3 \times 10^{9}$, leading to an upper limit $\delta ( M ) < \sqrt{\mu} \sim 10^{-2}$ over the mass range $10^{3} < M / \Msun < 10^{12}$. This limit was first given in Reference \cite{Carr:1993aq}, based on a result in Reference \cite{1991MNRAS.248...52B}, but the limit on $\mu$ is now much stronger. There is also a $y$ distortion for $3 \times 10^{9} < t /{\rm s}< 3 \times 10^{12}$.

This argument gives a very strong constraint on $f( M )$ in the range $10^{3} < M / \Msun < 10^{12}$ \cite{Kohri:2014lza} but the assumption that the fluctuations are Gaussian may be incorrect. For example, Nakama {\it et al.}~\cite{Nakama:2016kfq} have proposed a ``patch'' model, in which the relationship between the background inhomogeneities and the overdensity in the tiny fraction of volumes collapsing to PBHs is modified, so that the $\mu$-distortion constraint becomes much weaker. Recently, Nakama {\it et al.}~\cite{Nakama:2017xvq} have used a phenomenological description of non-Gaussianity to calculate the $\mu$-distortion constraints on $f( M )$, using the current FIRAS limit and the projected upper limit from PIXIE \cite{Abitbol:2017vwa}. However, one would need huge non-Gaussianity to avoid the constraints in the mass range of $10^{6}\,\Msun < M < 10^{10}\,\Msun$. Another way out is to assume that the PBHs are initially smaller than the lower limit but undergo substantial accretion between the $\mu$-distortion era and the time of matter-radiation equality.
The $\mu$ constraint also implies a maximum mass for the PBHs which provide the dark matter if there is some theoretical limit on the steepness of the power spectrum~\cite{Byrnes:2018txb}.

\subsection{Gravitational-Wave Constraints}
\label{sec:Gravitational--Wave-Constraints}

\noindent Interest in PBHs has intensified recently because of the detection of gravitational waves from coalescing black-hole binaries by LIGO/Virgo \cite{Abbott:2016blz, Abbott:2016nmj, Abbott:2016nhf, LIGOScientific:2018mvr}. Even if these are not of primordial origin, the observations place important constraints on the number of PBHs. Indeed, the LIGO data had already placed weak constraints on such scenarios a decade ago \cite{Abbott:2006zx}. More recent LIGO/Virgo searches find no compact binary systems with component masses in the range $0.2$ -- $1.0\,\Msun$ \cite{Abbott:2018oah}. Neither black holes nor neutron stars are expected to form through normal stellar evolution below $1\,\Msun$ and one can infer $f < 0.3$ for $M < 0.2\,\Msun$ and $f < 0.05$ for $M < 1\,\Msun$. A similar search from the second LIGO/Virgo run \cite{Authors:2019qbw} found constraints on the mergers of $0.2\,\Msun$ and $1.0\,\Msun$ binaries corresponding to at most $16\%$ or $2\%$ of the dark matter, respectively. The possibility that the LIGO/Virgo events relate to PBHs is discussed further in Section \ref{sec:LIGO/Virgo}.

A population of massive PBHs would be expected to generate a gravitational-wave background (GWB) \cite{1980A&A....89....6C} and this would be especially interesting if there were a population of binary black holes coalescing at the present epoch due to gravitational-radiation losses. Conversely, the non-observation of a GWB gives constraints on the fraction of dark matter in PBHs. As shown by Raidal {\it et al.}~\cite{Raidal:2017mfl}, even the early LIGO results gave strong limits in the range $0.5$ -- $30\,\Msun$ and this limit is shown in Figure \ref{fig:contraints-large}. A similar result was obtained by Wang {\it et al.}~\cite{Wang:2016ana}. This constraint has been updated in more recent work, using both LIGO/Virgo data ~\cite{Raidal:2018bbj, Vaskonen:2019jpv} and pulsar-timing observations \cite{Chen:2019xse}. Bartolo {\it et al.}~\cite{Bartolo:2019zvb} calculate the anisotropies and non-Gaussianity of such a stochastic GWB and conclude that PBHs could not provide all the dark matter if these were large.

A different type of gravitational-wave constraint on $f( M )$ arises because of the large second-order tensor perturbations generated by the scalar perturbations which produce the PBHs~\cite{Saito:2008jc}. The associated frequency was originally given as $10^{-8}\,( M / 10^{3}\,\Msun)\,{\rm Hz}$ but this estimate contained a numerical error \cite{Saito:2009jt} and was later reduced by a factor of $10^{3}$ \cite{Bugaev:2009zh}. The limit on $f( M )$ just relates to the amplitude of the density fluctuations at the horizon epoch and is of order $10^{-52}$. This effect has subsequently been studied by several other authors \cite{Assadullahi:2009jc, Bugaev:2010bb} and limits from LIGO/Virgo and the Big Bang Observer (BBO) could potentially cover the mass range down to $10^{20}\,$g. Conversely, one can use PBH limits to constrain a background of primordial gravitational waves \cite{Nakama:2015nea, Nakama:2016enz}. 

The robustness of the LIGO/Virgo bounds on $\Ocal( 10 )\,\Msun$ PBHs depends on the accuracy with which the formation of PBH binaries in the early Universe can be described. Ballesteros {\it et al.}~\cite{Ballesteros:2018swv} revisit the standard estimate of the merger rate, focusing on the spatial distribution of nearest neighbours and the expected initial PBH clustering. They confirm the robustness of the previous results in the case of a narrow mass function, which constrains the PBH fraction of dark matter to be $f \sim 0.001$ -- $0.01$. 

K{\"u}hnel {\it et al.}~\cite{Kuhnel:2018mlr} investigate GW production by PBHs in the mass range $10^{-13}$ -- $1\,\Msun$ orbiting a supermassive black hole. While an individual object would be undetectable, the extended stochastic emission from a large number of such objects might be detectable. In particular, LISA could detect the extended emission from objects orbiting Sgr\.${\rm A}^{\!*}$ at the centre of the Milky Way if a dark-matter spike, analogous to the WIMP-spike predicted by Gondolo and Silk \cite{Gondolo:1999ef}, forms there.

\subsection{Interesting Mass Windows and Extended Mass Functions}
\label{sec:Interesting-Mass-Windows-and-Extended-Mass-Functions}

\noindent Figure \ref{fig:contraints-large} shows that there are four mass windows (A,\.B,\.C,\.D) in which PBHs could have an ``appreciable'' density, which we somewhat arbitrarily take to mean $f > 0.1$, although this does not mean there is positive evidence for this. The cleanest window would seem to be A and many of the earlier constraints in this mass range have now been removed. Window C has received most attention, because of the LIGO/Virgo results, but it is challenging to put all the dark matter there because of the large number of constraints in this mass range.

A special comment is required about window D (\ie~the mass range $10^{14} < M / \Msun < 10^{18}$), since this has been almost completely neglected in previous literature. Obviously such stupendously large black holes (which we term ``SLABs'') could not provide the dark matter in galactic halos, since they are too large to fit inside them (\ie~they violate the galactic incredulity limit). However, they might provide an {\it intergalactic} dark-matter component and the lack of constraints in this mass range may just reflect the fact that nobody has considered this possibility. While this proposal might seem exotic, we know there are black holes with masses up to nearly $10^{11}\,\Msun$ in galactic nuclei \cite{Shemmer:2004ph}, so it is conceivable that SLABs could represent the high-mass tail of such a population. Although PBHs are unlikely to be this large at formation, we saw in Section \ref{sec:Accretion-Constraint} that they might increase their mass enormously before galaxy formation through accretion, so they could certainly {\it seed} SLABs. This possibility has motivated the study of such objects in Reference \cite{Carr:2020erq} and this includes an update of the accretion limit mentioned in Section \ref{sec:Accretion-Constraint} and a limit associated with WIMP annihilation, to be discussed in Section \ref{sec:Combined-Primordial-Black-Hole-and-Particle-Dark-Matter}. These limits are not shown in Figure \ref{fig:contraints-large} since that is just a summary of {\it previous} literature.

The constraints shown in Figure \ref{fig:contraints-large} assume that the PBH mass function is quasi-monochromatic (\ie~with a width $\Delta M \sim M$). This is unrealistic and in most scenarios one would expect the mass function to be extended, possibly stretching over several decades of mass. A detailed assessment of this problem requires a knowledge of the expected PBH mass fraction, $f_{\rm exp}( M )$, and the maximum fraction allowed by the monochromatic constraint, $f_{\rm max}( M )$. However, one cannot just plot $f_{\rm exp}( M )$ for a given model in Figure \ref{fig:contraints-large} and infer that the model is allowed because it does not intersect $f_{\rm max}( M )$. This problem is quite challenging and several different approaches have been suggested.

One approach is to assume that each constraint can be treated as a sequence of flat constraints by breaking it up into narrow mass bins ~\cite{Carr:2016drx} but this is a complicated procedure and has been criticised by Green \cite{Green:2016xgy}. A more elegant approach, similar to Green's, was proposed in Reference \cite{Carr:2017jsz} and also used in Reference \cite{Kuhnel:2017pwq}. In this, one introduces the function
\begin{equation}
	\psi( M )
		\propto
					M\,\frac{ \drm n }{ \drm M }
					\,,
\end{equation}
normalised so that the \emph{total} fraction of the dark matter in PBHs is
\begin{equation}
	f_{\rm PBH}
		\equiv
					\frac{ \Omega_{\rm PBH} }{ \Omega_{\rm CDM} }
		=
					\int_{M_{\rm min}}^{M_{\rm max}}\drm M\;\psi( M )
					\, .
\end{equation}
The mass function is specified by the mean and variance of the $\log M$ distribution:
\begin{equation}
	\log M_{\crm}
		\equiv
					\langle \log M \rangle^{}_{\psi}
					\, ,
	\quad
	\sigma^{2}
		\equiv
					\langle
						\log^{2} M
					\rangle^{}_{\psi}
					-
					\langle
						\log M
					\rangle_{\psi}^{2}
					\, ,
\end{equation}
where
\begin{equation}
	\langle X \rangle^{}_{\psi}
		\equiv
					f_{\rm PBH}^{-1}\,
					\int \drm M\;\psi( M )\,X( M )
					\, .
\end{equation}
Two parameters should generally suffice \emph{locally} (\ie~close to a peak), since these just correspond to the first two terms in a Taylor expansion. An astrophysical observable $A[ \psi( M ) ]$ depending on the PBH abundance can generally be expanded as
\begin{equation}
	A[ \psi( M ) ]
		=
					A_{0}
					+
					\int \drm M\;\psi( M )\,K_{1}( M )
					+
					\int \drm M_{1}\,\drm M_{2}\;
					\psi( M_{1} )\,\psi( M_{2} )\,K_{2}( M_{1},\,M_{2} )
					+
					\ldots
					\; ,
					\label{eq:A-observable}
\end{equation}
where $A_{0}$ is the background contribution and the functions $K_{j}$ depend on the details of the underlying physics and the nature of the observation. If PBHs with different mass contribute independently to the observable, only the first two terms in Equation \eqref{eq:A-observable} need be considered. If a measurement puts an upper bound on the observable,
\begin{equation}
	A[ \psi( M ) ]
		\leq
					A_{\rm exp}
					\, ,
					\label{eq:Aexp}
\end{equation}
then for a monochromatic mass function with $M = M_{\crm}$ we have
\begin{equation}
	\psi_{\mathrm{mon}}( M )
		\equiv
					f_{\rm PBH}( M_{\crm} )\,
					\delta( M - M_{\crm} )
					\, .
\end{equation}
The maximum allowed fraction of dark matter in the PBHs is then
\begin{equation}
	f_{\rm PBH}( M_{\crm} )
		\leq
					\frac{ A_{\rm exp} - A_{0} }
					{ K_{1}( M_{\crm} ) }
		\equiv
					f_{\rm max}( M_{\crm} )
					\, .
					\label{eq:f-max}
\end{equation}
Combining Equations \eqref{eq:A-observable}--\eqref{eq:f-max} then yields
\begin{equation}
	\int \drm M\;
	\frac{ \psi( M ) }
	{ f_{\rm max}( M )}
		\leq
					1\,.
					\label{eq:general-constraint}
\end{equation}
Once $f_{\rm max}$ is known, it is possible to apply Equation \eqref{eq:general-constraint} for an arbitrary mass function to obtain the constraints equivalent to those for a monochromatic mass function. One first integrates Equation \eqref{eq:general-constraint} over the mass range ($M_{1},\,M_{2}$) for which the constraint applies, assuming a particular function $\psi ( M;\,f_{\rm PBH},\,M_{\crm},\,\sigma )$\,. Once $M_{1}$ and $M_{2}$ are specified, this constrains $f_{\rm PBH}$ as a function of $M_{\crm}$ and $\sigma$. This procedure must be implemented separately for each observable and each mass function. 

In Reference \cite{Carr:2017jsz} this method is applied for various expected PBH mass functions, while Reference \cite{Kuhnel:2017pwq} performs a comprehensive analysis for the case in which the PBHs cover the mass range $10^{-18}$ -- $10^{4}\,\Msun$. Generally the allowed mass range for fixed $f_{\rm PBH}$ decreases with increasing width $\sigma$, thus ruling out the possibility of evading the constraints by simply extending the mass function. However, we stress that the situation could be more complicated than we have assumed above, with more than two parameters being required to describe the PBH mass function. For example, Hasegawa {\it et al.}~\cite{Hasegawa:2017jtk} have proposed an inflationary scenario in the minimally supersymmetric standard model which generates both intermediate-mass PBHs to explain the LIGO/Virgo detections and lunar-mass PBHs to explain the dark matter. Section~\ref{sec:Unified-Primordial-Black-Hole-Scenario} considers a scenario in which the PBH mass function has four peaks, each associated with a particular cosmological conundrum.

\section{Claimed Signatures}
\label{sec:Claimed-Signatures}

\noindent Most of the PBH literature has focussed on constraints on their contribution to the dark matter, as reviewed in the last Section. However, a number of papers have claimed positive evidence for them, the PBH masses required have been claimed to span this range over 16 orders of magnitude, from $10^{-10}\,\Msun$ to $10^{6}\,\Msun$. In particular, Reference \cite{Carr:2019kxo} summarises seven current observational conundra which may be explained by PBHs. The first three are associated with lensing effects: 
	(1) microlensing events towards the Galactic bulge 
		generated by planetary-mass objects \cite{Niikura:2019kqi} 
		which are much more frequent than expected
		for free-floating planets; 
	(2) microlensing of quasars \cite{Mediavilla:2017bok}, 
		including ones that are so misaligned with the lensing galaxy 
		that the probability of lensing by a star is very low; 
	(3) the unexpectedly high number of microlensing events towards the Galactic bulge 
		by dark objects in the `mass gap' between $2$ and $5\,\Msun$ 
		\cite{Wyrzykowski:2019jyg}, 
		where stellar evolution models fail to form black holes \cite{Brown:1999ax}. 
		The next three are associated with accretion and dynamical effects: 
	(4) unexplained correlations in the source-subtracted X-ray and 
		cosmic infrared background fluctuations \cite{2013ApJ...769...68C};
	(5) the non-observation of ultra-faint dwarf galaxies (UFDGs) 
		below the critical radius associated with dynamical disruption
		by PBHs \cite{Clesse:2017bsw}; 
	(6) the unexplained correlation between the masses of galaxies and 
		their central SMBHs. The final one is associated with gravitational-wave effects: 
	(7) the observed mass and spin 
		distributions for the coalescing black holes found by LIGO/Virgo \cite{LIGOScientific:2018mvr}. 
There are additional observational problems which Silk has argued may be solved by PBHs in the intermediate mass range \cite{Silk:2017yai}. We now discuss this evidence in more detail, discussing the conundra in order of increasing mass. In Section \ref{sec:Unified-Primordial-Black-Hole-Scenario} we discuss how these conundra may have a unified explanation in the scenario proposed in Reference \cite{Carr:2019kxo}.

\subsection{Lensing}
\label{sec:Lensing}

\noindent Observations of M31 by Niikura {\it et al.}~\cite{Niikura:2017zjd} with the HSC/Subaru telescope have identified a single candidate microlensing event with mass in the range range $10^{-10} < M < 10^{-6}\,\Msun$. Kusenko {\it et al.}~\cite{Kusenko:2020pcg} have argued that nucleation of false vacuum bubbles during inflation could produce PBHs with this mass. Niikura {\it et al.}~also claim that data from the five-year OGLE survey of 2622 microlensing events in the Galactic bulge \cite{Niikura:2019kqi} have revealed six ultra-short ones attributable to planetary-mass objects between $10^{-6}$ and $10^{-4}\,\Msun$. These would contribute about $1\%$ of the CDM, much more than expected for free-floating planets \cite{2019A&A...624A.120V}, and compatible with the bump associated with the electro-weak phase transition in the best-fit PBH mass function of Reference \cite{Carr:2019kxo}.
 
The MACHO collaboration originally reported $17$ LMC microlensing events and claimed that these were consistent with compact objects of $M \sim 0.5\,\Msun$, compatible with PBHs formed at the QCD phase transition~\cite{Alcock:2000ph}. Although they concluded that such objects could contribute only $20\%$ of the halo mass, the origin of these events is still a mystery and this limit is subject to several caveats. Calcino {\it et al.}~\cite{Calcino:2018mwh} argue that the usual semi-isothermal sphere for our halo is no longer consistent with the Milky Way rotation curve. When the uncertainties in the shape of the halo are taken into account, they claim that the LMC microlensing constraints weaken for $M \sim 10\,\Msun$ but tighten at lower masses. Hawkins~\cite{Hawkins:2015uja} makes a similar point, arguing that low-mass Galactic halo models would relax the constraints and allow $100\%$ of the dark matter to be solar-mass PBHs. Several authors have claimed that PBHs could form in tight clusters, giving a local overdensity well in excess of that provided by the halo concentration alone \cite{Dokuchaev:2004kr, Chisholm:2005vm}, and that this increased overdensity may remove the microlensing constraint at $M \sim 1$ -- $10\,\Msun$ altogether, especially if the PBHs have a wide mass distribution.

OGLE has detected around $60$ long-duration microlensing events in the Galactic bulge, of which around $20$ have GAIA parallax measurements. This finding breaks the mass-distance degeneracy and implies that these events were probably generated by black holes \cite{Wyrzykowski:2019jyg}. The event distribution implies a mass function peaking between $0.8$ and $5\,\Msun$, which overlaps with the gap from $2$ to $5\,\Msun$ in which black holes are not expected to form as the endpoint of stellar evolution \cite{Brown:1999ax}. This implication is also consistent with the peak originating from the reduction of pressure at the QCD epoch \cite{Carr:2019kxo}.

Hawkins \cite{1993Natur.366..242H} originally claimed evidence for a critical density of Jupiter-mass PBHs from observations of quasar microlensing. However, his later analysis yielded a lower density (dark matter rather than critical) and a mass of around $1 \,\Msun$ \cite{Hawkins:2006xj}. Mediavilla {\it et al.}~\cite{Mediavilla:2017bok} have also found evidence for quasar microlensing, this indicating that $20$\% of the total mass is in compact objects in the mass range $0.05$ -- $0.45\,\Msun $. These events might be explained by intervening stars but in several cases the stellar region of the lensing galaxy is not aligned with the quasar, which suggests a different population of subsolar halo objects. Hawkins \cite{Hawkins:2020zie} has also argued that some quasar images are best explained as microlensing by PBHs distributed along the lines of sight to the quasars. The best-fit PBH mass function of Reference \cite{Carr:2019kxo} is consistent with these findings and requires $\fPBH \simeq 0.05$ in this mass range. 

Recently Vedantham {\it et al.}~\cite{Vedantham:2017kyb} have detected long-term radio variability in the light-curves of some active galactic nuclei (AGN). This is associated with a pair of strongly skewed peaks in the radio flux density and is observed over a broad frequency range. They propose that this arises from gravitational millilensing of relativistically moving features in the AGN jets, these features crossing the lensing caustics created by $10^{3}$ -- $10^{6}\,\Msun$ subhalo condensates or black holes located within intervening galaxies.

\subsection{Dynamical}
\label{sec:Dynamical}

\noindent Lacey and Ostriker once argued that the observed puffing of the Galactic disc could be due to black holes of around $10^{6}\,\Msun$ \cite{1985ApJ...299..633L}, older stars being heated more than younger ones. They claimed that this could explain the scaling of the velocity dispersion with age and the relative velocity dispersions in the radial, azimuthal and vertical directions, as well as the existence of a high-velocity tail of stars \cite{1985AA...149..408I}. However, later measurements gave different velocity dispersions for older stars~\cite{1985ApJ...294..674C, 1987ASIC..207..229S, 1990AA...236...95G} and it is now thought that heating by a combination of spiral density waves and giant molecular clouds may better fit the data \cite{1991dodg.conf..257L}. 

If there were an appreciable number of PBHs in galactic halos, CDM-dominated UFDGs would be dynamically unstable if they were smaller than some critical radius which also depends on the mass of any central black hole. The non-detection of galaxies smaller than $r_{\crm} \sim 10$ -- $20\,$parsecs, despite their magnitude being above the detection limit, suggests the presence of compact halo objects in the solar-mass range. Recent $N$-body simulations \cite{Boldrini:2019isx} suggest that this mechanism works for PBHs of $25$ -- $100\,M_{\odot}$ providing they provide at least 1\% of the dark matter. Moreover, rapid accretion in the densest PBH haloes could explain the extreme UFDG mass-to-light ratios observed \cite{Clesse:2017bsw}. On the other hand, this model may not be supported by a recent analysis of Stegmann {\it et al.}~\cite{Stegmann:2019wyz}.

Fuller {\it et al.}~\cite{Fuller:2017uyd} show that some $r$-process elements can be produced by the interaction of PBHs with neutron stars if those in the mass range $10^{-14}$ -- $10^{-8}\,\Msun$ have $f > 0.01$. When a PBH is captured by a rotating millisecond neutron star, the resulting spin-up ejects $\sim 0.1$ -- $0.5\,\Msun$ of relatively cold neutron-rich material. This can also produce a kilonova-type afterglow and a fast radio burst. Abramowicz and Bejger \cite{Abramowicz:2017zbp} argue that collisions of neutron stars with PBHs of mass $10^{23}\,$g may explain the millisecond durations and large luminosities of fast radio bursts. 
 
As discussed in Section \ref{sec:Dynamical-Constraints}, sufficiently large PBHs could generate cosmic structures through the `seed' or `Poisson' effect \cite{Carr:2018rid}, the mass binding at redshift $z_{\Brm}$ being 
\begin{align}
	\bar{M}
		\approx
					\begin{cases}
						4000\.M\.z_{\Brm}^{-1}
							& ( {\rm seed} )
							\\[1mm]
						10^{7}\.f\.M\.z_{\Brm}^{-2}
							& ( {\rm Poisson} )
							\, .
						\end{cases}
					\label{eq:bind}
\end{align}
Having $f = 1$ requires $M < 10^{3}\,\Msun$ and so the Poisson effect could only bind a scale $\bar{M} < 10^{10}\.z_{\Brm}^{-2}\,\Msun$, which is necessarily subgalactic. However, this would still allow the dwarf galaxies to form earlier than in the standard scenario, which would have interesting observational consequences \cite{Silk:2017yai}. Having $f \ll 1$ allows the seed effect to be important and raises the possibility that the $10^{6}$ -- $10^{10}\,\Msun$ black holes in AGN are primordial in origin and {\it generate} the galaxies. For example, most quasars contain $10^{8}\,\Msun$ black holes, so it is interesting that this suffices to bind a region of mass $10^{11}\Msun$ at the epoch of galaxy formation. It is sometimes argued that the success of the BBN scenario requires PBHs to form before $1\,$s, corresponding to a limit $M < 10^{5}\,\Msun$. However, the fraction of the Universe in PBHs at time $t$ is only $10^{-6}\.\sqrt{ t / {\srm}\,}$, so the effect on BBN should be small. Furthermore, the softening of the pressure at $e^+e^-$ annihilation at $10\,$s naturally produces a peak at $10^{6}\,\Msun$. As discussed in Section \ref{sec:Accretion-Constraint}, such large PBHs would inevitably increase their mass through accretion, so this raises the question of whether a $10^{6}\,\Msun$ PBH would naturally grow to $10^{8}\,\Msun$, this entailing a corresponding increase in the value of $f$.

\subsection{X-Ray/Infrared Background}
\label{sec:X--Ray/Infrared-Background}

\noindent As shown by Kashlinsky and his collaborators~\cite{2013ApJ...769...68C, 2005Natur.438...45K, 2018RvMP...90b5006K, Kashlinsky:2016sdv}, the spatial coherence of the X-ray and infrared source-subtracted backgrounds suggests that black holes are required. Although these need not be primordial, the level of the infrared background suggests an overabundance of high-redshift haloes and this could be explained by the Poisson effect discussed above if a significant fraction of the CDM comprises solar-mass PBHs. In these haloes, a few stars form and emit infrared radiation, while PBHs emit X-rays due to accretion. It is challenging to find other scenarios that naturally produce such features.

\subsection{LIGO/Virgo}
\label{sec:LIGO/Virgo}

\noindent It has long been appreciated that a key signature of PBHs would the gravitational waves generated by either their formation \cite{1980A&A....89....6C}, although these would be hard to detect because of redshift effects, or their coalescences if the PBHs form binaries. Indeed, the detection of coalescing binary black holes was first discussed in Reference \cite{1984MNRAS.207..585B} in the context of Population III black holes and later in References \cite{Nakamura:1997sm, Ioka:1998gf} in the context of PBHs. However, the precise formation epoch of the holes is not crucial since the coalescence occurs much later. In either case, the black holes would be expected to cluster inside galactic halos and so the detection of the gravitational waves would provide a probe of the halo distribution \cite{Inoue:2003di}. 

The suggestion that the dark matter could comprise PBHs has attracted much attention in recent years as a result of the LIGO/Virgo detections \cite{Abbott:2016blz, Abbott:2016nmj}. To date, $10$ events have been observed with component masses in the range $8$ -- $51\,\Msun$. After the first detection, Bird {\it et al.}~\cite{Bird:2016dcv} claimed that the expected merger was compatible with the range $9$ -- $240\,{\rm Gpc}^{-3}\,{\rm years}^{-1}$ obtained by the LIGO analysis and this was supported by other studies~\cite{Clesse:2016vqa, Blinnikov:2016bxu}. On the other hand, Sasaki {\it et al.}~\cite{Sasaki:2016jop} argued that the lower limit on the merger rate would be in tension with the CMB distortion constraints if the PBHs provided all the dark matter, although one might avoid these constraints if the LIGO/Virgo back holes derive from the accretion and merger of smaller PBHs \cite{Clesse:2016vqa}. Note that most of the observed coalesced black holes have effective spins compatible with zero. Although the statistical significance of this result is low \cite{Fernandez:2019kyb}, it goes against a stellar binary origin \cite{Gerosa:2018wbw} but is a prediction of the PBH scenario \cite{Garcia-Bellido:2017fdg}. 

If the PBHs have an extended mass function, their density should peak at a lower-mass signal than the coalescence signal. For example, $\fPBH^{\rm tot} = 1$ but $\fPBH( M ) \sim 0.01$ in the range $10$ -- $100\,\Msun$ for the mass distribution of Reference \cite{Carr:2019kxo}. Raidal {\it et al.}~\cite{Raidal:2017mfl} have studied the production and merging of PBH binaries for an extended mass function and possible PBH clustering (cf.~\cite{Dolgov:2020xzo}). They show that PBHs can explain the LIGO/Virgo events without violating any current constraints if they have a lognormal mass function. Subsequent work~\cite{Raidal:2018bbj, Vaskonen:2019jpv} has studied the formation and disruption of PBH binaries in more detail, using both analytical and numerical calculations for a general mass function. If PBHs make up just 10\% of the dark matter, the analytic estimates are reliable and indicate that the constraint from the observed LIGO/Virgo rate is strongest in the mass range $2$ -- $160\,\Msun$, albeit weakened because of the suppression of mergers. Their general conclusion is that the LIGO/Virgo events can result from the mergers of PBHs but that such objects cannot provide all the dark matter unless the PBHs have an extended mass function. 

Ali-Ha{\"i}moud {\it et al.}~\cite{Ali-Haimoud:2017rtz} have computed the probability distribution of orbital parameters for PBH binaries. Their analytic estimates indicate that the tidal field of halos and interactions with other PBHs, as well as dynamical friction by unbound standard dark-matter particles, do not provide a significant torque on PBH binaries. They also calculate the binary merger rate from gravitational capture in present-day halos. If binaries formed in the early Universe survive to the present time, as expected, they dominate the total PBH merger rate. Moreover, this merger rate would be above the current LIGO upper limit unless $f( M ) < 0.01$ for $10$ -- $300\,\Msun$ PBHs. 

One of the mass ranges in which PBHs could provide the dark matter is around $10^{-12}\,\Msun$. If these PBHs are generated by enhanced scalar perturbations produced during inflation, their formation is inevitably accompanied by the generation of non-Gaussian gravitational waves with frequency peaked in the mHz range (the maximum sensitivity of LISA). Bartolo {\it et al.}~\cite{Bartolo:2018rku, Bartolo:2018evs} have studied whether LISA will be able to detect not only the gravitational-wave (GW) power spectrum but also the non-Gaussian three-point GW correlator (\ie~the bispectrum). However, they conclude that the inclusion of propagation effects suppresses the bispectrum. If PBHs with masses of $10^{20}$ -- $10^{22}\,$g are the dark matter, the corresponding GWs will be detectable by LISA, irrespective of the value of $f_{\rm NL}$, and this has also been stressed by Cai {\it et al.}~\cite{Cai:2018dig}.

\subsection{Arguments for Intermediate-Mass Primordial Black Holes}
\label{sec:Arguments-for-Intermediate--Mass-Primordial-Black-Holes}

\noindent Silk has argued that intermediate-mass PBHs (IMPBHs) could be ubiquitous in early dwarf galaxies, being mostly passive today but active in their gas-rich past~\cite{Silk:2017yai}. This would be allowed by current AGN observations \cite{2013ARAA..51..511K, 2016ApJ...831..203P, Baldassare:2016cox} and early feedback from IMPBHs could provide a unified explanation for many dwarf galaxy anomalies. Besides providing a phase of early galaxy formation and seeds for SMBHs at high $z$ (discussed above), they could:
	(1) suppress the number of luminous dwarfs; 
	(2) generate cores in dwarfs by dynamical heating; 
	(3) resolve the ``too big to fail'' problem; 
	(4) create bulgeless disks; 
	(5) form ultra-faint dwarfs and ultra-diffuse galaxies; 
	(6) reduce the baryon fraction in Milky-Way-type galaxies; 
	(7) explain ultra-luminous X-ray sources in the outskirts of galaxies;
	(8) trigger star formation in dwarfs via AGN. 
As we will see in Section \ref{sec:Thermal-History-of-the-Universe}, IMPBH production could be naturally triggered by the thermal history of the Universe~\cite{Carr:2019kxo}. This would lead to other observational signatures: they would generate extreme-mass-ratio inspiral merger events detectable by LISA; they would tidally disrupt white dwarfs much more rapidly than main-sequence stars, leading to luminous flares and short time-scale nuclear transients \cite{Law-Smith:2017agr}; they would induce microlensing of extended radio sources \cite{Inoue:2003hq, Inoue:2013ey}.
\vs{2mm}

\section{Unified Primordial Black Hole Scenario}
\label{sec:Unified-Primordial-Black-Hole-Scenario}

\noindent In this Section we describe a particular scenario in which PBHs naturally form with an extended mass function and provide a unified explanation of some of the conundra discussed above. The scenario is discussed in detail Reference \cite{Carr:2019kxo} and based on the idea that the thermal history of the Universe leads to dips in the sound-speed and therefore enhanced PBH formation at scales corresponding to the electroweak phase transition ($10^{-6}\,\Msun$), the QCD phase transition ($1\,\Msun$), the pion-plateau ($10\,\Msun$) and $e^{+}\.e^{-}$ annihilation ($10^{6}\,\Msun$). This scenario requires that most of the dark matter is in PBHs formed at the QCD peak and is marginally consistent with the constraints discussed in Section \ref{sec:Constraints-and-Caveats}, even though this suggests that the QCD window cannot provide all the dark matter for a monochromatic PBH mass function.

\subsection{Thermal History of the Universe}
\label{sec:Thermal-History-of-the-Universe}

\noindent Reheating at the end of inflation fills the Universe with radiation. In the standard model, it remains dominated by relativistic particles with an energy density decreasing as the fourth power of the temperature. As time increases, the number of relativistic degrees of freedom remains constant until around $200\,$GeV, when the temperature of the Universe falls to the mass thresholds of the Standard Model particles. The first particle to become non-relativistic is the top quark at $172\,$GeV, followed by the Higgs boson at $125\,$GeV, the $Z$ boson at $92\,$GeV and the $W$ boson at $81\,$GeV. At the QCD transition at around $200\,$MeV, protons, neutrons and pions condense out of the free light quarks and gluons. A little later the pions become non-relativistic and then the muons, with $e^{+}e^{-}$ annihilation and neutrino decoupling occur at around $1\,$MeV.

\begin{figure}
	\centering
	\vs{-3mm}
	\includegraphics[width = 0.50\textwidth]{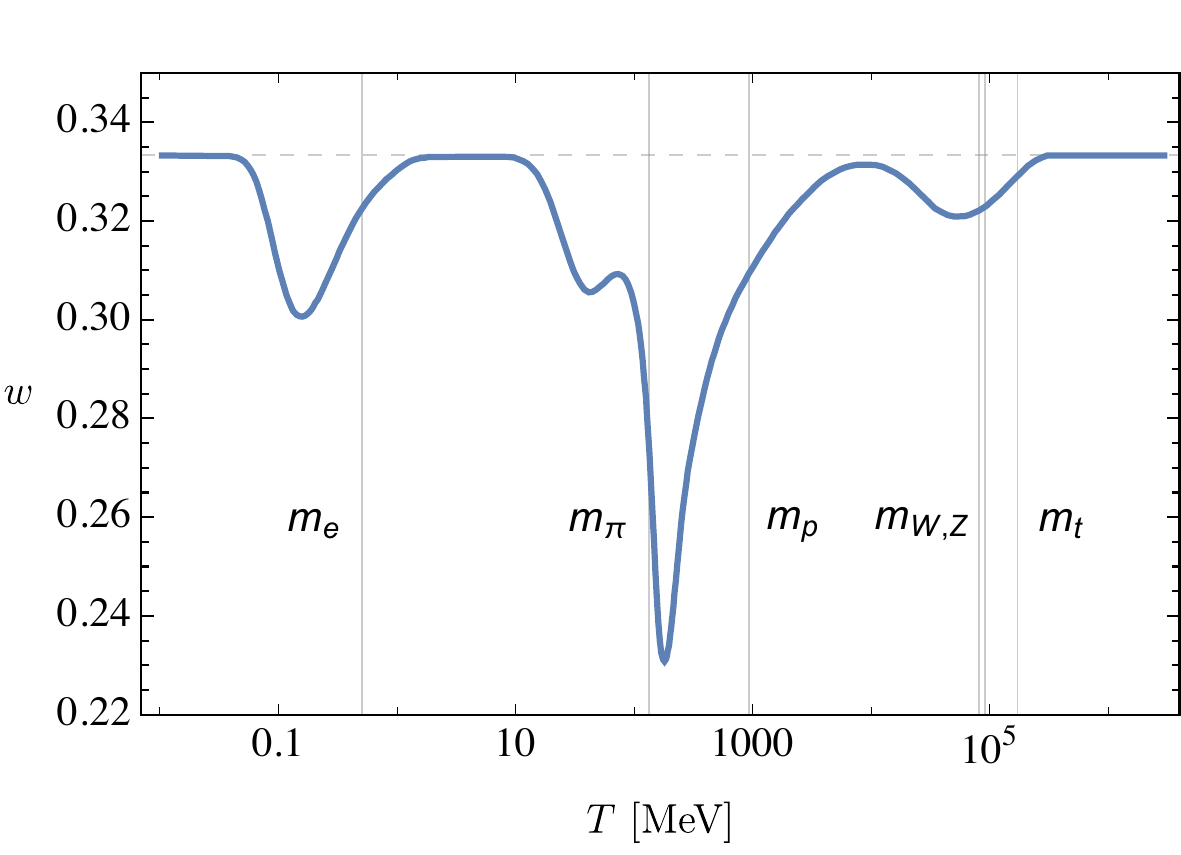}
	\caption{
			Equation-of-state parameter $w$ 
			as a function of temperature $T$, 
			from Reference \cite{Carr:2019kxo}.
			The grey vertical lines correspond to the 
			masses of the electron, 
			pion, proton/neutron, $W / Z$ bosons and 
			top quark, respectively.
			The grey dashed horizontal lines 
			correspond to $g_{*} = 100$ 
			and $w = 1 / 3$.}
	\label{fig:g-and-w-of-T}
\end{figure}

Whenever the number of relativistic degrees of freedom suddenly drops, it changes the effective equation of state parameter $w$. As shown Figure \ref{fig:g-and-w-of-T}, there are thus four periods in the thermal history of the Universe when $w$ decreases. After each of these, $w$ resumes its relativistic value of $1 / 3$ but because the threshold $\delta_{\crm}$ is sensitive to the equation-of-state parameter $w( T )$, the sudden drop modifies the probability of gravitational collapse of any large curvature fluctuations. This results in pronounced features in the PBH mass function even for a uniform power spectrum. If the PBHs form from Gaussian inhomogeneities with root-mean-square amplitude $\delta_{\rm rms}$, then Equation \eqref{eq:beta} implies that the fraction of horizon patches undergoing collapse to PBHs when the temperature of the Universe is $T$ is \cite{Carr:1975qj}
\begin{align}
	\beta( M )
		\approx
				{\rm Erfc}\!
				\left[
					\frac{
						\delta_{\crm}
						\big(
							w[ T( M ) ]
						\big) }
					{ \sqrt{2}\,\delta_{\rm rms}( M )}
				\right]
				,
				\label{eq:beta( T )}
\end{align}
where the value $\delta_{\crm}$ comes from Reference \cite{Musco:2012au} and the temperature is related to the PBH mass by 
\begin{align}
	T
		\approx
				200\,\sqrt{\Msun / M\,}\;\MeV
				\, .
\end{align}
Thus $\beta( M )$ is exponentially sensitive to $w( M )$ and the present CDM fraction for PBHs of mass $M$ is 
\begin{align}
	\fPBH( M ) 
		\equiv
				\frac{ 1 }{ \rho_{\rm CDM} }
				\frac{ \drm\.\rho_{\rm PBH}( M ) }
				{ \drm \ln M }
		\approx
				2.4\;\beta( M )\.
				\sqrt{\frac{ M_{\rm eq} }{ M }\,}
				\, ,
				\label{eq:fPBH}
\end{align}
where $M_{\rm eq} = 2.8 \times 10^{17}\,\Msun$ is the horizon mass at matter-radiation equality and the numerical factor is $2\.( 1 + \Omega_{\Brm} / \Omega_{\rm CDM} )$ with $\Omega_{\rm CDM} = 0.245$ and $\Omega_{\Brm} = 0.0456$ \cite{Aghanim:2018eyx}. This is equivalent Equation \eqref{eq:beta2}.

\begin{figure}
	\vs{-3mm}
	\centering
	\includegraphics[width = 0.5 \textwidth]{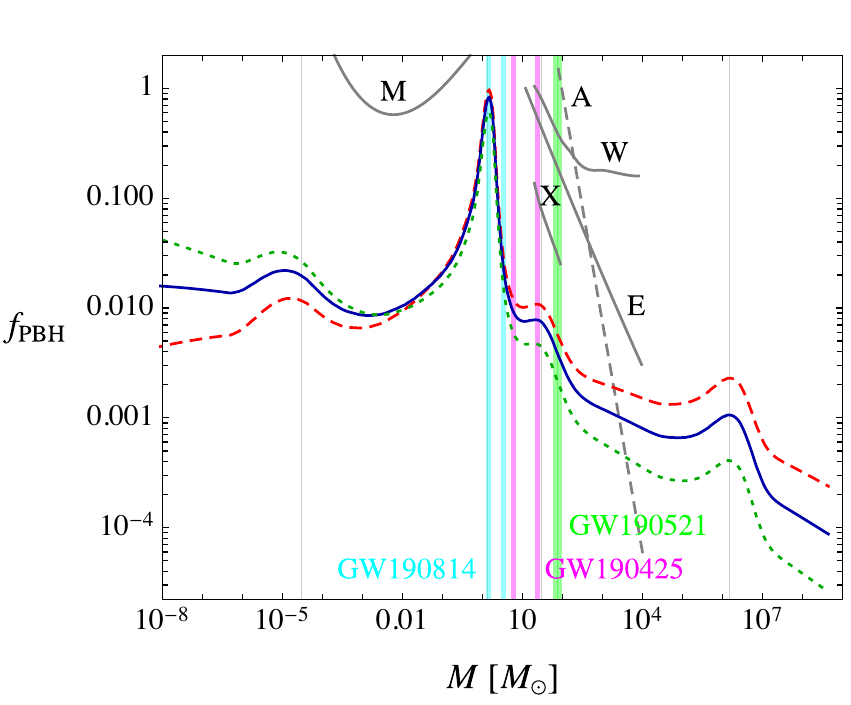}
	\caption{The mass spectrum of PBHs
			with spectral index 
			$n_{\srm} = 0.965$ (red, dashed),
			$0.97$ (blue, solid),
			$0.975$ (green, dotted), from Reference \cite{Carr:2019kxo}.
			The grey vertical lines corresponds to the electroweak and 
			QCD phase transitions 
			and $e^{+}e^{-}$ annihilation. 
			Also shown are the constraints associated with 
			microlensing (M), 
			wide-binaries (W), 
			accretion (A), 
			Eridanus (E) and 
			X-ray observations (X).
			The vertical lines correspond to the gravitational-wave events 
			GW190425 \cite{Abbott:2020uma}, GW190814 \cite{Abbott:2020khf} 
			and GW190521 \cite{Abbott:2020mjq, Abbott:2020tfl}.
			}
	\label{fig:fPBH}
\end{figure}

There are many inflationary models and they predict a variety of shapes for $\delta_{\rm rms}( M )$. Some of them produce an extended plateau or dome-like feature in the power spectrum. For example, this applies for two-field models like hybrid inflation \cite{Clesse:2015wea} and even some single-field models like Higgs inflation \cite{Ezquiaga:2017fvi, Garcia-Bellido:2017mdw}, although this may not apply for the minimal Higgs model \cite{Bezrukov:2017dyv}. Instead of focussing on any specific scenario, Reference \cite{Carr:2019kxo} assumes a quasi-scale-invariant spectrum,
\begin{align}
	\delta_{\rm rms}( M )
		=
				A
				\left(
					\frac{ M }{ \Msun }
				\right)^{\!(1 - n_{\srm}) / 4}
				\; ,
				\label{eq:delta-power-law}
\end{align}
where the spectral index $n_{\srm}$ and amplitude $A$ are treated as free phenomenological parameters. This could represent any spectrum with a broad peak, such as might be generically produced by a second phase of slow-roll inflation. The amplitude is chosen to be $A = 0.0661$ for $n_{\srm} = 0.97$ in order to get an integrated abundance $\fPBH^{\rm tot} = 1$. The ratio of the PBH mass and the horizon mass at re-entry is denoted by $\gamma$ and we assume $\gamma = 0.8$, following References \cite{Carr:2019hud, Garcia-Bellido:2019vlf}. The resulting mass function is represented in Figure \ref{fig:fPBH}, together with the relevant constraints from Section \ref{sec:Constraints-and-Caveats}. It exhibits a dominant peak at $M \simeq 2\,\Msun$ and three additional bumps at $10^{-5}\,\Msun$, $30\,\Msun$ and $10^{6}\,\Msun$. 

\begin{figure}
	\centering 
	\includegraphics[width = 0.55 \textwidth]{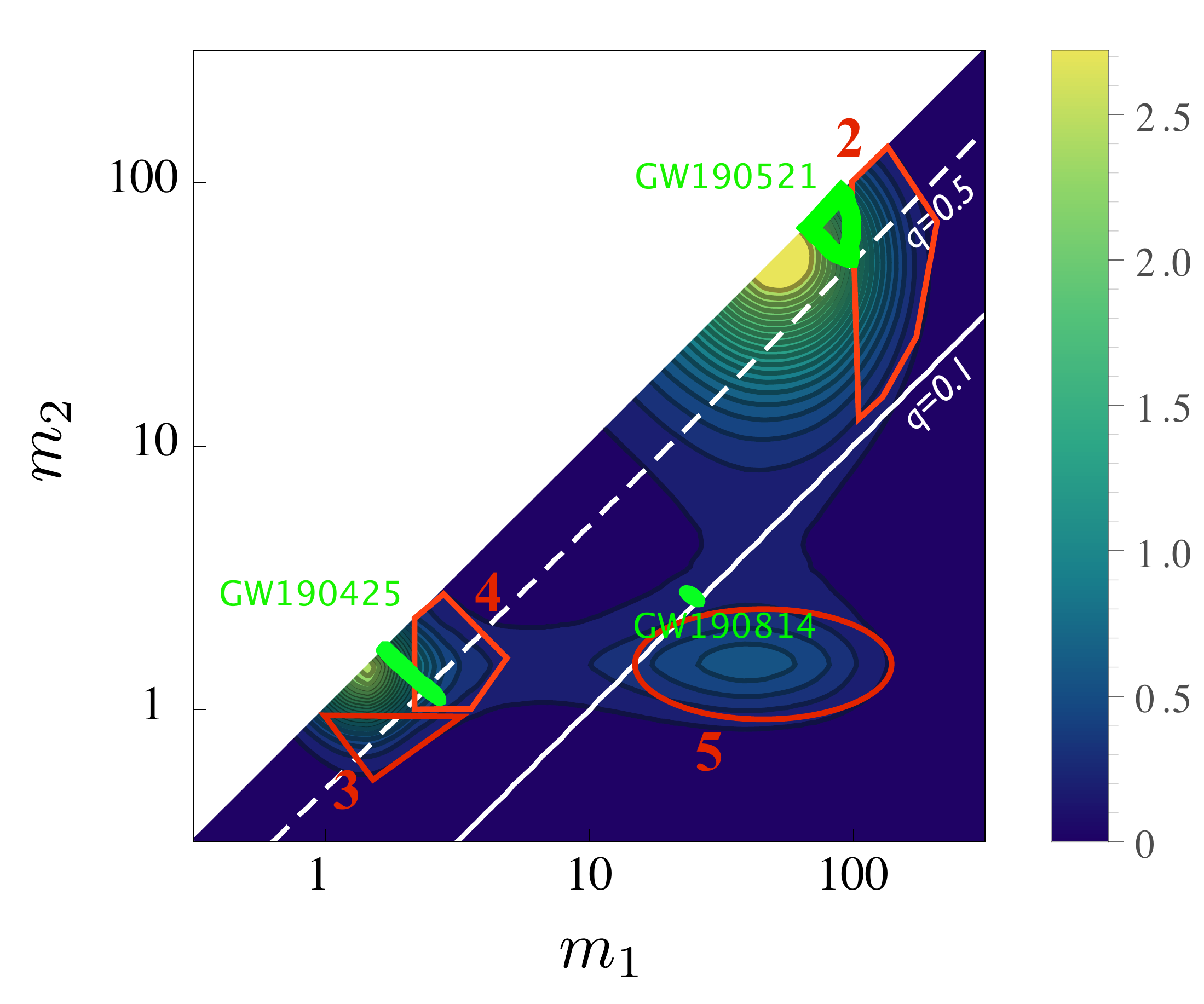}
	\caption{
			Expected probability distribution of PBH mergers
			with masses $m_{1}$ and $m_{2}$ (in solar units), 
			assuming a PBH mass function with $n_{\srm} = 0.97$ 
			and the LIGO sensitivity for the O2 run.
			The solid and dashed white lines correspond to mass ratios 
			$q \equiv m_{2} / m_{1}$ of $0.1$ and $0.5$, respectively.
			(1)~corresponds to the peak for neutron-star mergers without electromagnetic counterparts.
			The mergers of stellar black holes are not expected within the red-bouned regions, 
			which are:
			(2)~events with one black hole above $100\,\Msun$;
			(3)~mergers of subsolar objects,
				which might be taken to be neutron stars, with
				objects at peak of black-hole distribution;
			(4)~mergers of objects in mass gap;
			(5)~a subdominant population of mergers with low mass ratios.
			The colour bar indicates the probability of detection.
			The green lines indicate the gravitational-wave events 
			GW190425, GW190814 and GW190521, 
			these lying in regions (2), (4) and (5), respectively.
			Figure adapted from Reference \cite{Carr:2019kxo}.}
	\label{fig:Rate-LIGO-events}
\end{figure}

The discussion in Section \ref{sec:Claimed-Signatures} has already indicated how these bumps relate to various observational conundra and further details can be found in Reference \cite{Carr:2019kxo}. Here we emphasise two further features. First, observations of the mass ratios in coalescing binaries provide an important probe of the scenario, the distribution predicted in the unified model being shown in Figure \ref{fig:Rate-LIGO-events}. The regions in areas outlined by red lines are not occupied by stellar black-hole mergers in the standard scenario and the distinctive prediction is the merger of objects with $1\,\Msun$ and $10\,\Msun$, corresponding to region (5). 
Recently the LIGO/Virgo collaboration has reported the  detection of three  events, all of which (remarkably) fall within the predicted regions. Two of the them (GW190425 and GW190814) involve mergers with one component in the $2 - 5 \, M_{\odot}$ mass gap \cite{Abbott:2020uma,Abbott:2020khf} and populate regions 4 or 5 of Fig.~\ref{fig:Rate-LIGO-events}. The first could be a merger of PBHs at the ``proton'' peak , while the second corresponds to the ``pion'' plateau, as also argued in Reference~\cite{Jedamzik:2020omx}. The third event (GW190521) involves a merger  with at least one component in the pair-instability mass gap~\cite{Abbott:2020tfl,Abbott:2020mjq} and populates region 2 of Fig.~\ref{fig:Rate-LIGO-events}. Second, for a given PBH mass distribution, one can calculate the number of supermassive PBHs for each halo. It is found that there is one $10^{8}\,\Msun$ PBH per $10^{12}\,\Msun$ halo, with $10$ times as many smaller ones for $n_{\srm} \approx 0.97$ and $\fPBH^{\rm tot} \simeq 1$. If one assumes a standard Press-Schechter halo mass function and identifies the PBH mass that has the same number density, one obtains the relation $M_{\hrm} \approx M_{\rm PBH} / \fPBH$, in agreement with observations \cite{Kruijssen:2013cna}.
 
\subsection{Resolving the Fine-Tuning Problem}
\label{sec:Resolving-the-Fine--Tuning-Problem}

\noindent The origin of the baryon asymmetry of the Universe (BAU) and the nature of dark matter are two of the most challenging problems in cosmology. The usual assumption is that high-energy physics generates the baryon asymmetry everywhere simultaneously via out-of-equilibrium particle decays or a first-order phase transition at very early times. Garc{\'i}a-Bellido {\it et al.}~\cite{Garcia-Bellido:2019vlf} propose a scenario in which the gravitational collapse of large inhomogeneities at the QCD epoch (invoked above) can resolve both these problems. The collapse to a PBH is induced by fluctuations of a light spectator scalar field and accompanied by the violent expulsion of surrounding material, which might be regarded as a sort of ``primordial supernova". This provides the ingredients for efficient baryogenesis around the collapsing regions, with the baryons subsequently propagating to the rest of the Universe, and naturally explains why the observed BAU is of order the PBH collapse fraction
and why the baryons and dark matter have comparable densities. 

We now discuss this proposal in more detail. The gravitational collapse of the mass within the QCD Hubble horizon can be extremely violent~\cite{Musco:2012au} with particles being driven out as a relativistically expanding shock-wave and acquiring energies a thousand times their rest mass from the gravitational potential energy released by the collapse. Such high density {\it hot spots} provide the out-of-equilibrium conditions required to generate a baryon asymmetry \cite{Sakharov:1967dj} through the well-known electroweak sphaleron transitions responsible for Higgs windings around the electroweak vacuum~\cite{Asaka:2003vt}. In this process, the charge-parity (CP) symmetry violation of the Standard Model suffices to generate a local baryon-to-photon ratio of order one. The hot spots are separated by many horizon scales but the outgoing baryons propagate away from the hot spots at the speed of light and become homogeneously distributed well before BBN. The large initial local baryon asymmetry is thus diluted to the tiny observed global BAU.

\begin{figure}
	\vs{-14mm}
	\centering 
	\includegraphics[width = 0.8 \textwidth]{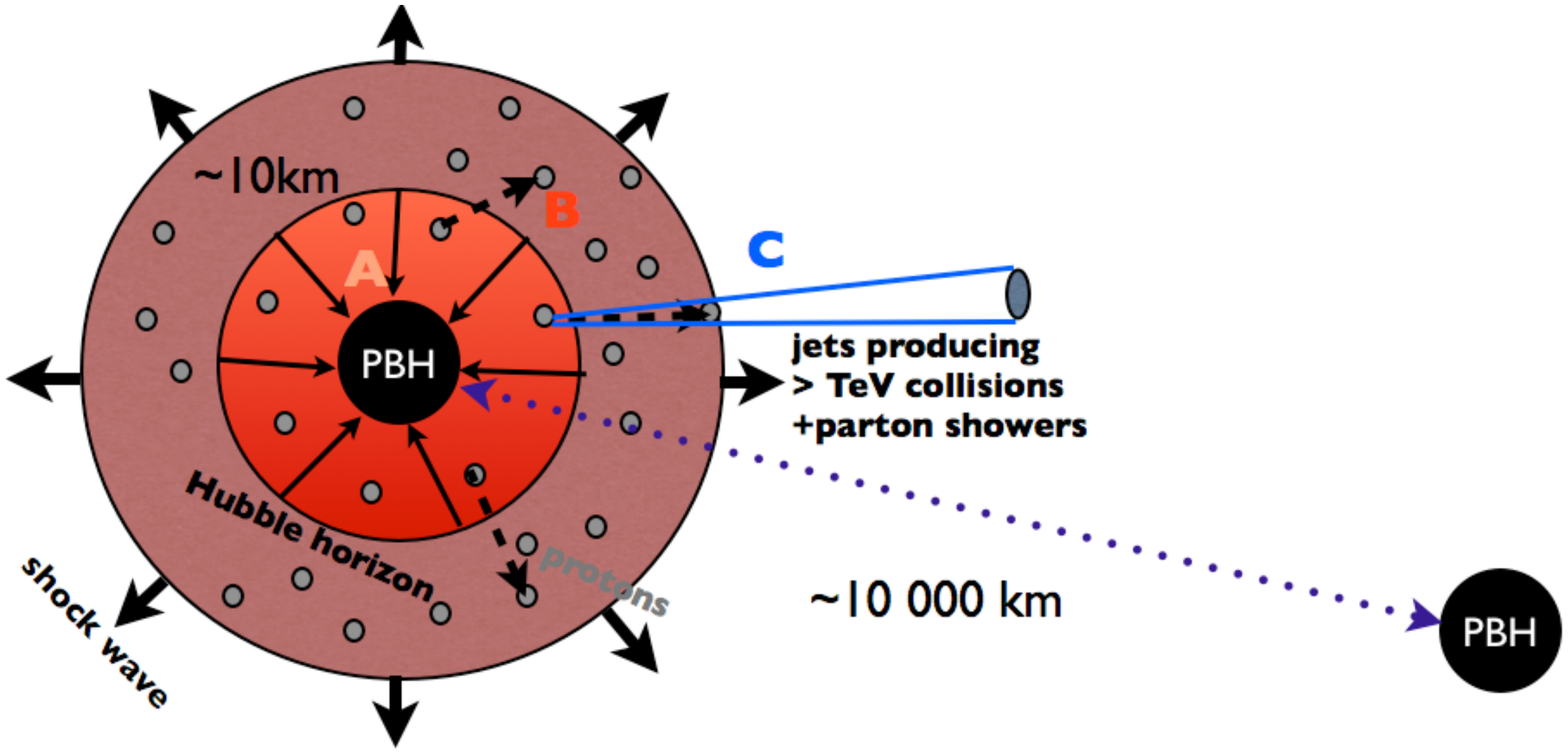}
	\vs{-18mm}
	\caption{
			Qualitative representation of the three 
			steps in the discussed scenario. 
			(A) Gravitational collapse to a PBH of 
				the curvature fluctuation at horizon 
				re-entry. 
			(B) Sphaleron transition in hot spot 
				around the PBH, 
				producing $\eta \sim \Ocal( 1 )$ 
				locally through electroweak baryogenesis. 
			(C) Propagation of baryons to rest of 
				Universe through jets, 
				resulting in the observed BAU with 
				$\eta \sim 10^{-9}$.
				Figure adapted from Reference \cite{Garcia-Bellido:2019vlf}.}
	\label{fig:BAU}
\end{figure}

The energy available for hot spot electroweak baryogenesis can be estimated as follows. Energy conservation implies that the change in kinetic energy due to the collapse of matter within the Hubble radius to the Schwarzschild radius of the PBH is
\begin{align}
	\Delta K
		\simeq
				\left(
					\frac{ 1 }{ \gamma }
					-
					1
				\right)\mspace{-1mu}
				M_{\Hrm}
		=
				\left(
					\frac{ 1 - \gamma }{ \gamma^{2} }
				\right)\mspace{-1mu}
				\MPBH
				\, .
\end{align}
The energy acquired per proton in the expanding shell is $E_{0} = \Delta K / ( n_{\prm}\,\Delta V)$, where $\Delta V = ( 1 - \gamma^{3} )\.V_{\Hrm}$ is the difference between the Hubble and PBH volumes, so $E_{0}$ scales as $( \gamma + \gamma^{2} +\gamma^{3} )^{-1}$. For a PBH formed at $T\approx \Lambda_{\rm QCD} \approx 140\,\MeV$, the effective temperature is $T_{\rm eff} = 2\.E_{0} / 3 \approx 5\,$TeV, which is well above the sphaleron barrier and induces a CP violation parameter $\dCP( T ) \sim 10^{-5} (T / 20 \GeV)^{-12}$~\cite{Shaposhnikov:2000sja}. The production of baryons can be very efficient, giving $\nB \ga n_{\gamma}$ or $\eta \ga 1$ locally. The scenario is depicted qualitatively in Figure \ref{fig:BAU}.

This proposal naturally links the PBH abundance to the baryon abundance and the BAU to the PBH collapse fraction ($\eta \sim \beta$). The spectator field mechanism for producing the required curvature fluctuations also avoids the need for a fine-tuned peak in the power spectrum, which has long been considered a major drawback of PBH scenarios. One still needs fine-tuning of the mean field value to produce the observed values of $\eta$ and $\beta$ (\ie~{$\sim 10^{-9}$}). However, the stochasticity of the field during inflation ensures that Hubble volumes exist with all possible field values and this means that one can explain the fine-tuning by invoking a single anthropic selection argument. The argument is discussed in Reference \cite{Carr:2019hud} and depends on the fact that only a small fraction of patches will have the PBH and baryon abundance required for galaxies to form. In most patches the field is too far from the slow-roll region to produce either PBHs or baryons, leading to radiation universes without any dark matter or matter-antimatter asymmetry. In other (much rarer) patches, PBHs are produced too copiously, leading to rapid accretion of most of the baryons, as might have happened in ultra-faint dwarf galaxies. This anthropic selection effect may therefore explain the observed values of $\eta$ and $\beta$. 

\section{Primordial Black Holes versus Particle Dark Matter}
\label{sec:Primordial-Black-Hole-versus-Particle-Dark-Matter}

\noindent Presumably most particle physicists would prefer the dark matter to be elementary particles rather than PBHs, although there is still no direct evidence for this. However, even if this transpires to be the case, we have seen that PBHs could still play an important cosmological r{\^o}le, so we must distinguish between PBHs providing {\it some} dark matter and {\it all} of it. This also applies for the particle candidates. Nobody would now argue that neutrinos provide the dark matter but they still play a hugely important r{\^o}le in astrophysics. Therefore one should not necessarily regard PBHs and particles as rival candidates. Both could exist and we end by discussing two scenarios in this spirit. The first assumes that particles dominate the dark matter but that PBHs still provide an interesting interaction with them. The second involves the notion that evaporating black holes leave stable Planck-mass (or even sub-Planck-mass) relics, although such relics are in some sense more like particles than black holes.

\subsection{Combined Primordial Black Hole and Particle Dark Matter}
\label{sec:Combined-Primordial-Black-Hole-and-Particle-Dark-Matter}

\noindent If most of the dark matter is in the form of elementary particles, these will be accreted around any small admixture of PBHs. In the case of WIMPs, this can even happen during the radiation-dominated era, since Eroshenko \cite{Eroshenko:2016yve} has shown that a low-velocity subset will accumulate around PBHs as density spikes shortly after the WIMPs kinetically decouple from the background plasma. Their annihilation will give rise to bright $\gamma$-ray sources and comparison of the expected signal with Fermi-LAT data then severely constrains $\Omega_{\rm PBH}$ for $M > 10^{-8}\,\Msun$. These constraints are several orders of magnitude more stringent than other ones if one assumes a WIMP mass of $m_{\chi} \sim \Ocal( 100 )\,\GeV$ and the standard value of $\langle \sigma v \rangle^{}_{\Frm} = 3 \times 10^{-26}\,{\rm cm}\,\srm^{-1}$ for the velocity-averaged annihilation cross-section. Boucenna {\it et al.}~\cite{Boucenna:2017ghj} have investigated this scenario for a larger range of values for $\langle \sigma v \rangle$ and $m_{\chi}$ and reach similar conclusions. This could also affect the evolution of binary PBHs, with consequent implications for LIGO/Virgo observations~\cite{Kavanagh:2018ggo}.

Apart from the early formation of spikes around PBHs which are light enough to arise very early, WIMP accretion around heavier PBHs can also occur by secondary infall \cite{Bertschinger:1985pd}. This leads to a different halo profile, yielding a constraint $f_{\rm PBH} \lesssim \Ocal( 10^{-9} )$ for the same values of $\langle \sigma v \rangle$ and $m_{\chi}$. While Adamek {\it et al.}~\cite{Adamek:2019gns} have derived this limit for solar-mass PBHs, the argument can be extended to much bigger masses, even up to the values associated with stupendously large black holes \cite{Carr:2020erq}. The constraint at intermediate $M$ comes from the integrated effect of a population of such objects and is flat:
\begin{equation}
	f_{\rm PBH}
		<
					\frac{ 16 }{ 3 }\.
					\frac{ \Phi^{\rm Fermi}_{100\.\MeV}\,H_{0} }
				 	{ \rho_{\rm DM}\,\tilde{N}_{\gamma}( m_{\chi} ) }\!
					\left(
						\frac{ 2\.m_{\chi}^{4}\,t_{0}^{2} }
						{ \langle\sigma v \rangle\,\rho_{\rm eq} }
					\right)^{\!\!1 / 3}
					\, ,
					\label{eq:SLAP-constraint}
\end{equation}
where $\Phi^{\rm Fermi}_{100\.\MeV} = 6 \times 10^{-9}\,{\rm cm}^{-2}\,\srm^{-1}$ is the Fermi point-source sensitivity above the threshold energy $E_{\rm th} = 100\,$MeV, $t_{0}$ is the age of the Universe and $\rho_{\rm eq} = 2.1 \times 10^{-19}\,{\rm g\,cm^{-3}}$ is the density at $t_{\rm eq}$. $\tilde{N}_{\gamma}$ is the average number of photons produced,
\begin{equation}
	\tilde{N}_{\gamma}( m_{\chi} )
		\equiv
				\int_{E_{\rm th}}^{m_{\chi}}\!\d E\;
				\frac{ \d N_{\gamma} }{ \d E }
				\int_{0}^{\infty}\!\d z\;
				\frac{ H_{0} }
				{ H( z ) }\;
				\exp\!
				\left[
					-
					\tau(z,\.E)
				\right]
				\, ,
\end{equation}
where $\d N_{\gamma} / \d E$ is the number of $\gamma$-rays emitted from the annihilations occurring per unit time and energy. The optical depth $\tau$ is the result of processes such as photon-matter pair production, photon-photon scattering and photon-photon pair production~\cite{Cirelli:2009dv}. The limit \eqref{eq:SLAP-constraint} is derived in Reference \cite{Carr:2020erq} using the numerical packages from Reference \cite{Cirelli:2010xx} to obtain the optical depth and spectrum of by-products from WIMP annihilations.

\begin{figure}[tb]
	\vs{4mm}
	\includegraphics[width = 0.60\linewidth]{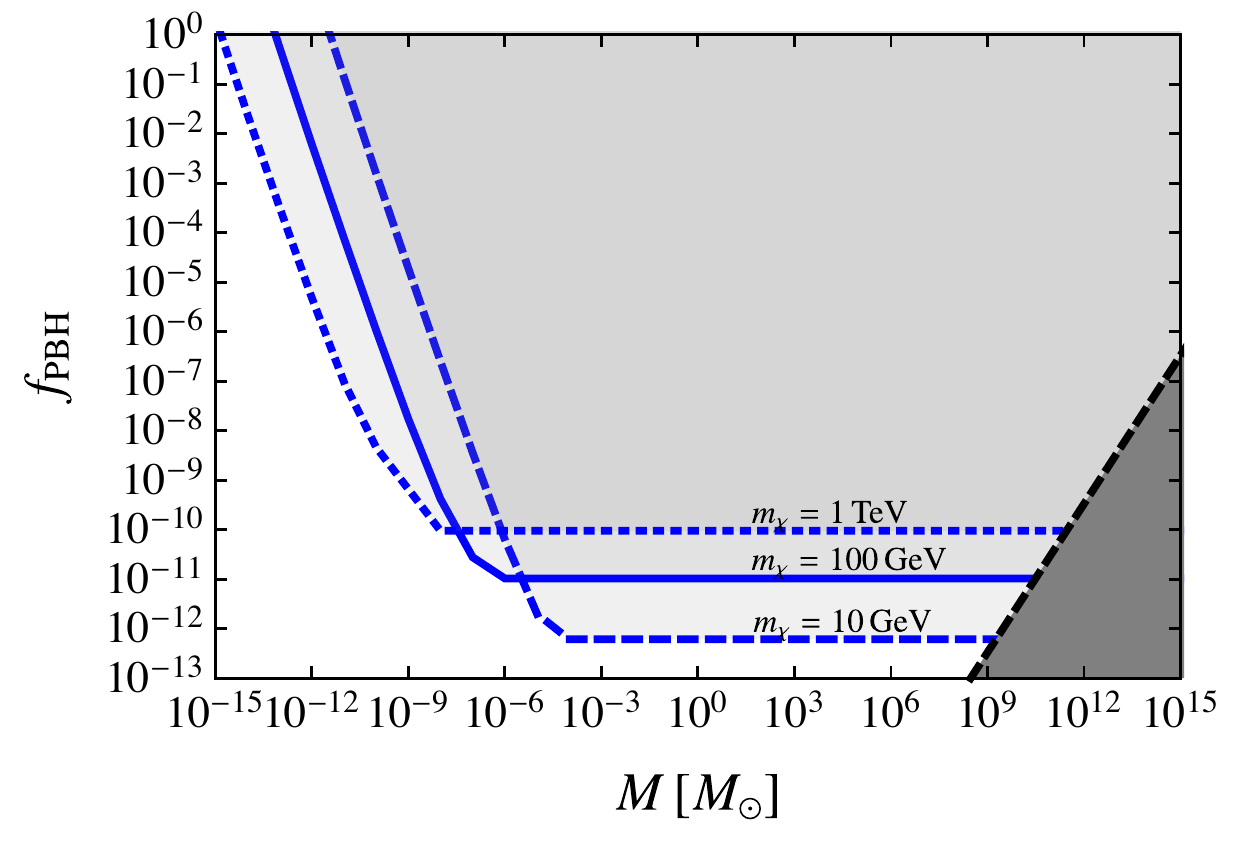} 
	\caption{Constraints on $f_{\rm PBH}$ as a function of PBH mass,
				for a WIMP mass of 
				$10\,$GeV (blue dashed line), 
				$100\,$GeV (blue solid line) and 
				$1\,$TeV (blue dotted line), 
				with $\langle \sigma v \rangle = 3 \times 10^{-26}\,$cm$^{3}\,$s$^{-1}$.
				Also shown is the incredulity limit (black dashed line), 
				corresponding to one PBH per horizon.
				Figure adapted from Reference \cite{Carr:2020erq}.}
	\label{fig:fPBHWIMP}
\end{figure}

Figure~\ref{fig:fPBHWIMP} shows constraints on $\fPBH$ for WIMP masses of $10\,$GeV, $100\,$GeV and $1\,$TeV. The falling part at low $M$ is associated with halos formed after dark-matter kinetic decoupling (when the kinetic energy of the WIMPs is important); the flat part is associated with halos formed by secondary infall at later times (when the kinetic energy can be neglected). No bound can be placed above the mass where the lines intersects the incredulity limit
\begin{align}
	M_{\rm IL}
		 = 
					2.5 \times 10^{10}
					\left(
						\frac{ m_{\chi} }{ 100\,\GeV }
					\right)^{\!1.11}
					\left(
						\frac{ \langle \sigma v \rangle}{ \langle \sigma v \rangle_{\Frm} }
					\right)^{\!\!-1 / 3}
					\Msun
					.
\end{align}
For axion-like particles or sterile neutrinos, there is a similar limit but from decays rather than annihilations. K{\"u}hnel \& Ohlsson \cite{Kuhnel:2018kwf} have derived bounds on the axion-like particle (ALP) and found that the detection prospects for ALP masses below $\Ocal( 1 )\,\keV$ and halos heavier than $10^{-5}\,\Msun$ are far better than for the pure ALP scenario. For sterile-neutrino halos, there are good detection prospects for X-ray, $\gamma$-ray and neutrino telescopes \cite{Kuhnel:2017ofn, Kuhnel:2020xrn}.

\subsection{Planck-Mass Relics}
\label{sec:Planck--Mass-Relics}

\noindent If PBH evaporations leave stable Planck-mass relics, these might also contribute to the dark matter. This was first pointed out by MacGibbon \cite{MacGibbon:1987my} and subsequently explored in the context of inflationary scenarios by several other authors~\cite{Carr:1994ar, Barrow:1992hq, Alexander:2007gj, Fujita:2014hha}. If the relics have a mass $\kappa\,M_{\rm Pl}$ and reheating occurs at a temperature $T_{\Rrm}$\., then the requirement that they have less than the critical density implies \cite{Carr:1994ar}
\begin{equation}
	\beta'( M )
		<
					2 \times 10^{-28}\,\kappa^{-1}\!
					\left(
						\frac{ M }{ M_{\rm Pl} }
					\right)^{\!3/2}
					\label{eq:betarelic}
\end{equation}
for the mass range
\vs{-3mm}
\begin{equation}
	\left(
		\frac{ T_{\mathrm{Pl}} }{ T_{\Rrm} }
	\right)^{\!2}
		<
					\frac{ M }
					{ M_{\rm Pl} }
		<
					10^{11}\,\kappa^{2 / 5}
					\, .
					\label{eq:relic}
\end{equation}
One would now require the density to be less than $\Omega_{\rm CDM} \approx 0.26$, which strengthens the original limit by a factor of $4$. The lower mass limit arises because PBHs generated before reheating are diluted exponentially. The upper mass limit arises because PBHs larger than this dominate the total density before they evaporate, in which case the final cosmological baryon-to-photon ratio is determined by the baryon-asymmetry associated with their emission. Limit \eqref{eq:betarelic} still applies even if there is no inflationary period but then extends all the way down to the Planck mass.

It is usually assumed that such relics would be undetectable apart from their gravitational effects. However, Lehmann {\it et al.}~\cite{Lehmann:2019zgt} have recently pointed out that they may carry electric charge, making them visible to terrestrial detectors. They evaluate constraints and detection prospects and show that this scenario, if not already ruled out by monopole searches, can be explored within the next decade with planned experiments. 

In some scenarios PBHs could leave stable relics whose mass is very different from the Planck mass. For example, if one maintains the Schwarzschild expression but adopts the Generalised Uncertainty Principle, in which $\Delta x \sim 1 / \Delta p + \alpha\.\Delta p$, then evaporation stops at a mass $\sqrt{\alpha}\,M_{\mathrm{Pl}\,}$ \cite{Chen:2002tu}. On the other hand, if one adopts the Black Hole Uncertainty Principle correspondence~\cite{Carr:2013mqa}, in which one has a unified expression for the Schwarzschild and Compton (SC) scales, $R_{\rm SC} = 2\.G M + \beta\.M_{\rm Pl}^{2} / M$, the mass can fall into the sub-Planckian regime in which $T \propto M$. In this case, evaporation stops when the black hole becomes cooler than the CMB at $M \sim 10^{-36}\,$g \cite{Carr:2015nqa}. One motivation for this correspondence is Dvali \& Gomez's framework for black holes as graviton Bose-Einstein condensates \cite{Dvali:2011aa}. K{\"u}hnel and Sandstad \cite{Kuhnel:2015qaa} have studied PBH formation in this context and argued that evaporation may stop because the gravitons are quantum-depleted much faster than the baryons or leptons, which are ``caught'' in the black-hole condensate. So at some point the balance of the strong and gravitational forces leads to stable relics. Recently, Dvali {\it et al.}~\cite{Dvali:2020wft} have shown that the decay of a black hole is substantially suppressed by its high capacity for memory storage, leading to another mechanism for long-lived or even stable relics.

\section{Conclusions}
\label{sec:Conclusion}

\noindent While the study of PBHs has been a minority interest for most of the last 50 years, they have become the focus of increasing attention recently. This is strikingly reflected in the annual publication rate on the topic, which has now risen to several hundred. While the evidence for PBHs is far from conclusive, there is a growing appreciation of their many potential r{\^o}les in cosmology and astrophysics. This is why we have stressed the possible evidence for PBHs in this review rather than just the constraints.

PBHs have been invoked for three main purposes: 
	(1) to explain the dark matter; 
	(2) to generate the observed LIGO/Virgo coalescences; 
	(3) to provide seeds for the SMBHs in galactic nuclei. 
The discussion in Section \ref{sec:Unified-Primordial-Black-Hole-Scenario} suggests that they could also explain several other observational conundra, as well as alleviating some of the well-known problems of the CDM scenario. So PBHs could play an important cosmological r{\^o}le even if most of the dark matter transpires to be elementary particles. 

As regards (1), there are only a few mass ranges in which PBHs could provide the dark matter. We have focused on the intermediate mass range $10\,\Msun < M < 10^{2}\,\Msun$, since this may be relevant to (2), but the sublunar range $10^{20}$ -- $10^{24}\,$g and the asteroid range $10^{16}$ -- $10^{17}\,$g have also been suggested. If the PBHs have a monochromatic mass function, the discussion in Section \ref{sec:Constraints-and-Caveats} suggests that only the lowest mass range is viable. However, the discussion in Section \ref{sec:Unified-Primordial-Black-Hole-Scenario} indicates that this conclusion may not apply if they have an extended mass function.

As regards (2), while the possibility that the LIGO/Virgo sources could be PBHs is acknowledged by the gravitational-wave community, this is not the mainstream view. It is therefore important to stress that the next LIGO/Virgo runs should be able to test and possibly eliminate the PBH proposal. Indeed, it is remarkable that the three recent events GW190425, GW190814 and GW190521 fall precisely within regions (2), (4) and (5) of Figure \ref{fig:Rate-LIGO-events}. In any case, (2) does not require the PBHs to provide {\it all} the dark matter. If the PBHs have an extended mass function, the mass where the density peaks would be less than the mass which dominates the gravitational-wave signal.

As regards (3), there is no reason in principle why the maximum mass of a PBH should not be in the supermassive range, in which case it is almost inevitable that they could seed SMBHs and perhaps even galaxies themselves. The main issue is whether there are enough PBHs to do so but this only requires them to have a very low cosmological density. While the mainstream assumption is that galaxies form first, with the SMBHs forming in their nuclei through dynamical processes, this is not certain.
 A crucial question concerns the growth of such large black holes and this applies whether or not they are primordial.

Section~\ref{sec:Unified-Primordial-Black-Hole-Scenario} has described a scenario in which PBHs form with a bumpy mass function as a result of naturally occurring dips in the sound-speed at various cosmological epochs, this naturally explaining explain many cosmological conundra. This scenario also suggests that the cosmological baryon asymmetry may be generated by PBH formation at the QCD epoch, this naturally explaining the fine-tuning in the collapse fraction. This is not the mainstream view for the origin of the baryon asymmetry and this proposal require further investigation but this is a first attempt to address the much-neglected PBH fine-tuning problem. The possibility that evaporating PBHs leave stable relics opens up some of the mass range below $10^{15}\,$g as a new world of compact dark-matter candidates waiting to be explored.

\acknowledgments

\noindent We are grateful to S{\'e}bastien Clesse, Juan Garc{\'i}a-Bellido, Kazunori Kohri, Marit Sandstad, Yuuiti Sendouda, Luca Visinelli and Jun-ichi Yokoyama, some of our joint work being reported in this review. We also thank our many other PBH collaborators. Helpful remarks from Yacine Ali Ha{\"i}moud, Chris Belczynski, Valerio de Luca, Sasha Dolgov, Xiao Fang, Gabriele Francolini, Cristiano Germani, Anne Green, Qing-Guo Huang, Ranjan Laha, Paulo Monteiro-Camacho, Shi Pi, Martti Raidal, Javier Rubio, Toni Riotto, Jakob Stegmann, Haijun Tian, Matteo Viel, Sai Wang and Miguel Zumalac{\'a}rregui are warmly acknowledged. B.C.~thanks the Research Center for the Early Universe (RESCEU) of University of Tokyo and F.K.~thanks the Oskar Klein Centre for Cosmoparticle Physics, Queen Mary University of London and the Delta Institute for Theoretical Physics for hospitality and support. F.K.~also acknowledges support from the Swedish Research Council through contract No.~638-2013-8993.

\setlength{\bibsep}{5pt}
\setstretch{1}
\bibliography{refs}

\end{document}